\documentclass[prb,aps,twocolumn,showpacs]{revtex4-1}
\usepackage{amsmath}
\usepackage{dcolumn}
\usepackage{amsfonts}
\usepackage{amssymb}
\usepackage{graphicx}

\begin{document}
\title{Topological insulators and fractional quantum Hall effect on the ruby lattice}
\date{\today}
\author{Xiang Hu}
\email{phyxiang@gmail.com}
\author{Mehdi Kargarian}
\author{Gregory A. Fiete}
\affiliation{Department of Physics, The University of Texas at Austin, Austin, Texas 78712, USA}
\begin{abstract}
We study a tight-binding model on the two-dimensional ruby lattice.
This lattice supports several types of first and second neighbor spin-dependent
hopping parameters in an $s$-band model that preserves time-reversal
symmetry.  We discuss the phase diagram of this model for various
values of the hopping parameters and filling fractions, and note an interesting competition
between spin-orbit terms that individually would drive the
system to a $Z_2$ topological insulating phase.  We also discuss a closely related spin-polarized model with only first and second neighbor hoppings and show that extremely flat bands with finite Chern numbers result, with a ratio of the band gap to the band width approximately 70.  Such flat bands are an ideal platform to realize a fractional quantum Hall effect at appropriate filling fractions.  The ruby lattice can be possibly engineered in optical lattices, and may open the door to studies of transitions between quantum spin liquids, topological insulators, and integer and fractional quantum Hall states.
\end{abstract}



\maketitle

\section{Introduction}

Topological phases of matter have received a great deal of attention recently.\cite{Wen,Volovik}  In part, this is motivated by the fractional quantum Hall effect and its possible role as a platform for topological quantum computation.\cite{Nayak:rmp08} However, there is also a more general interest in phases of quantum many-particle systems that can exhibit responses and other properties that are a consequence of global ({\em i.e.} topological) features that are not captured in a local order parameter.\cite{Levin:prl06,Eisert:rmp10,Amico:rmp08} A key example of such a phase is the time-reversal invariant topological insulator.\cite{Qi:pt10,Moore:nat10,Hasan:rmp10}  In contrast to most of the previously known systems that exhibited some topological property (the integer quantum Hall effect being a notable exception), topological insulators do not require electron-electron interactions, but they are stable to interactions of weak to moderate strength.\cite{Qi:pt10,Moore:nat10,Hasan:rmp10}  The weakly-interacting nature of topological insulators has enabled accurate predictions\cite{Bernevig:sci06,Fu:prb07,Teo:prb08,Zhang:np09} for a wide range of two and three dimensional systems,\cite{Chadov:nm10,Wang:prl11,Feng:prl11,Zhang:prl11,Xiao:prl10,Lin:prl10,Chen:prl10,Yan:epl10,Yan:prb10,Lin:cm10,Yan:11,Lin:nm10,Sun:prl10,Lin:prl10,Feng:prb10,Al-Sawai:prb10} and experiment has followed with confirming data in a large and rapidly growing number of instances.\cite{Konig:sci07,Roth:sci09,Hsieh:nat08,Hsieh:sci09,Xia:np09,Chen:sci09,Hsieh:nat09,Hsieh:prl09,Hor:prb09,Sato:prl10,Kuroda:prl10,Nishide:prb10}

\begin{figure}[t]
\includegraphics[width=0.75\linewidth]{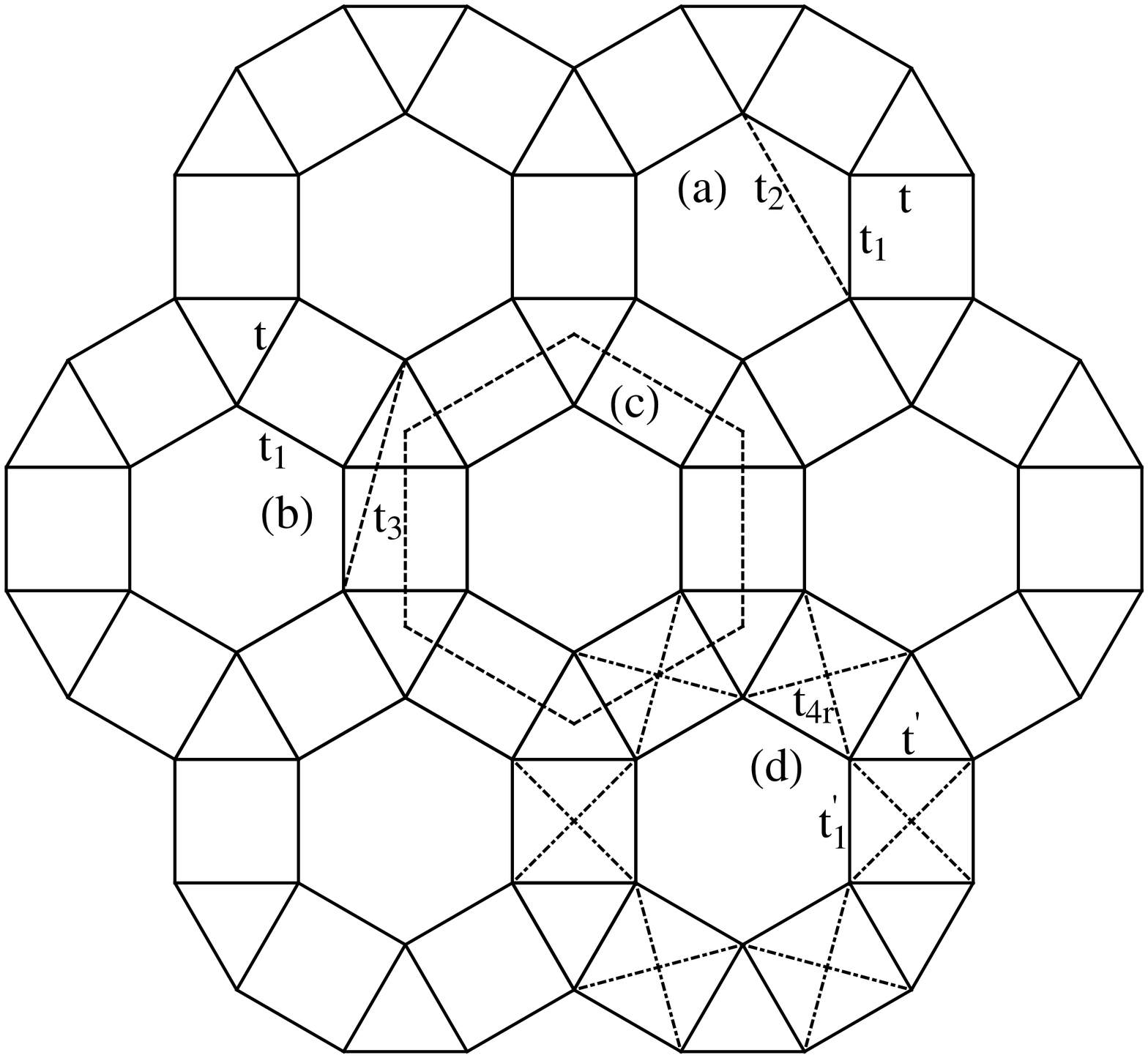}
\includegraphics[width=0.23\linewidth]{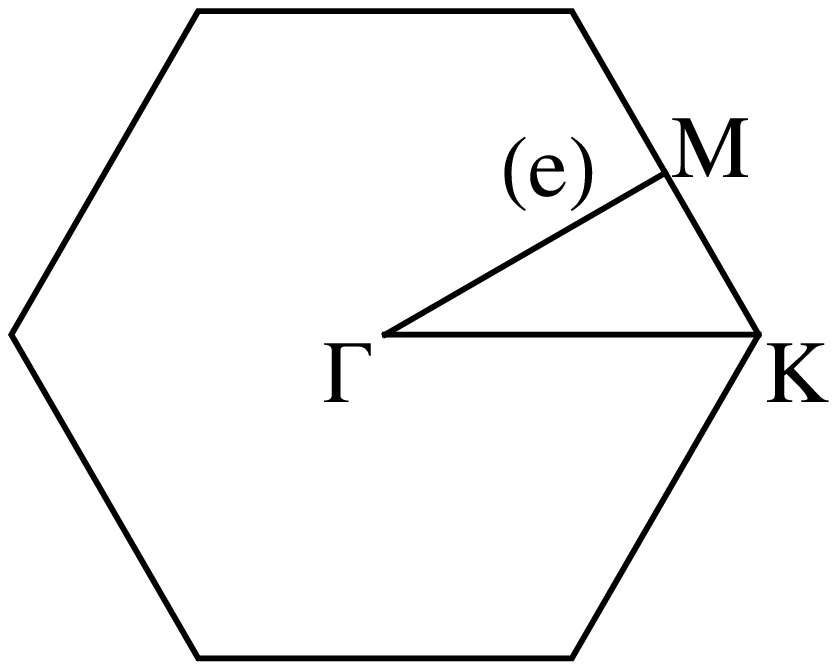}
\caption{Schematic of the ruby lattice and illustration of the nearest-neighbor hopping, $t,t_1$ (real) and $t',t_1'$ (complex), and the three types of ``second-neighbor" spin-orbit coupling or hopping indicated by the dashed or dot dashed lines, $t_2,t_3$ and $t_{4r}$. (a) The spin-orbit coupling strength $t_2$ within a hexagon. (b) The spin-orbit coupling strength $t_3$ within a pentagon composed of one triangle and one square.  The hoppings $t_2$ and $t_3$ are present on all such bonds of the type shown that are consistent with the symmetry of the lattice. (c) The unit cell of the ruby lattice. (d) Schematic of the hopping parameters used to obtain a flat band with a finite Chern number and $W/E_g \approx 70$. (e) The first Brillouin zone of the ruby lattice, with the high symmetry points $\Gamma$, $M$ and $K$.}\label{fig:ruby}
\end{figure}

The salient feature of topological insulators is that their boundaries possess a topologically protected ``metallic" state that is robust to disorder.\cite{Schnyder:prb08,Roy_2:prb09,Roy:prb09,Moore:prb07,Fu:prl07,Wu:prl06,Xu:prb06}  Both the two dimensional\cite{Strom:prl09,Huo:prl09,Xu:prb10,Teo:prb09,Law:prb10,Zyuzin:prb10,Tanaka:prl09,Maciejko:prl09} and the three dimensional\cite{Qi:prb08,Qi:sci09,Essin:prl09,Essin:prb10} boundaries have been shown to exhibit interesting responses to perturbations.  In this work we focus on a two dimensional tight-binding model with a single $s$-orbital on each lattice site.  We study the so-called ``ruby" lattice shown in Fig.\ref{fig:ruby}.  We find that it exhibits a complex phase diagram that includes topological insulators at a number of filling fractions.  This lattice has earlier played an important role in the study of topological order in spin models.\cite{Bombin:prb09,Kargarian:njp10}

Starting with the work of Kane and Mele\cite{Kane:prl05,Kane_2:prl05} on the honeycomb lattice (and key earlier precedents by Haldane\cite{Haldane:prl88} in a spinless version), such simple non-interacting lattice models have helped to develop our understanding of topological insulators.\cite{Bernevig:prl06}  Besides the honeycomb lattice,\cite{Kane:prl05,Kane_2:prl05} a number of other two dimensional lattices have been shown to support a topological insulator phase, including the decorated honeycomb lattice,\cite{Ruegg:prb10} the checkerboard lattice,\cite{Sun:prl09} the square-octagon lattice,\cite{Kargarian:prb10} the kagome lattice,\cite{Guo:prb09} and others.\cite{Weeks:prb10}  It is now well appreciated that such non-interacting lattice models possess topological features that commonly occur in interacting models {\em without spin-orbit coupling} at the mean-field level.\cite{Raghu:prl08,Sun:prl09,Wen:prb10,Zhang:prb09,Liu:prb10} Evidently, such simple lattice models contain rather rich physics.

An interesting related topic of study is the class of insulators with nearly flat bands that possess finite Chern numbers.   With a finite Chern number, a partial filling of the flat bands can lead to a fractional quantum Hall effect.\cite{Sheng11}  To date, only a few lattice models have been proposed which are expected to lead to a fractional quantum Hall effect.\cite{Sun10,Neupert11,Tang10,Wang11}  The relevant figure of merit in such models is the ratio of the band gap to the bandwidth of the flat band with a finite Chern number.  In the model we discuss in this paper, we find this ratio can be as high as 70, which is among the largest in the models reported in the literature thus far.

The remainder of the paper is organized as follows. In Sec.~\ref{sec:Hamiltonian} we introduce our tight-binding model, and discuss the basic features of the energy bands as a function of the hopping parameters. In Sec.~\ref{sec:phase}  we discuss the phase diagrams at different filling fractions, and in Sec.~\ref{sec:FQHE} we discuss a closely related spinless (spin polarized) model with nearly flat bands and finite Chern number.  Our main conclusions are given in Sec.~\ref{sec:conclusions}.

\section{Hamiltonian and Band Structure}
\label{sec:Hamiltonian}

We study the Hamiltonian
\begin{equation}
\label{eq:H}
H=H_0+H_{\rm SO},
\end{equation}
where
\begin{equation}
\label{eq:H0}
H_0=-t\sum_{i,j\in\triangle, \sigma}c_{i\sigma}^\dag c_{j\sigma}
    -t_1\sum_{\triangle\rightarrow\triangle,\sigma}c_{i\sigma}^\dag
    c_{j\sigma},
\end{equation}
and
\begin{equation}
\label{eq:H_SO}
H_{\rm SO}=it_2\!\!\!\!\sum_{\ll ij\gg, \alpha\beta}\!\!\!\!\nu_{ij}s^z_{\alpha\beta}
c_{i\alpha}^\dag c_{j\beta}+it_3\!\!\!\!\sum_{\ll ij\gg, \alpha\beta}\!\!\!\!\nu_{ij}
s^z_{\alpha\beta}c^\dagger_{i\alpha}c_{j\beta}
\end{equation}
on the ruby lattice shown in Fig.~\ref{fig:ruby}. Here $c_{i\sigma}^\dag/c_{i\sigma}$ is the creation/annihilation operator of an electron on site
$i$ with spin $\sigma$. As indicated in Fig.~\ref{fig:ruby}(a,b), $t$ and $t_1$ are real first-neighbor hopping parameters, and $t_2$, $t_3$ are real second-neighbor hoppings (these appear with the imaginary number $i$ in Eq.\eqref{eq:H_SO} making the total second-neighbor hopping purely imaginary and time-reversal symmetric). The quantity  $\nu_{ij}$ is equal to 1 if the electron makes a left turn on the lattice links during the second-neighbor hopping, and is equal to -1 if the electron make a right turn during that process. As is clear from Fig.\ref{fig:ruby}, the unit cell of the lattice contains six sites so six two-fold degenerate bands will result.  In addition to the real hopping parameters $t,t_1$ in \eqref{eq:H0}, symmetry also allows complex, spin-dependent nearest neighbor hopping with imaginary components $t',t_1'$, as shown in Fig.~\ref{fig:ruby}.  We will discuss terms of this type later in Sec.~\ref{sec:FQHE}.

The full Hamiltonian \eqref{eq:H} can be diagonalized by going to a momentum space representation,
\begin{equation}
H=\sum_{{\bf k}\sigma}\Psi^\dag_{{\bf k}\sigma}\tilde{H}_{{\bf k}\sigma}\Psi_{{\bf k}\sigma},
\end{equation}
where
$\Psi^\dag_{{\bf k}\sigma}=(c^\dag_{1{\bf k}\sigma},c^\dag_{2{\bf k}\sigma},
c^\dag_{3{\bf k}\sigma},c^\dag_{4{\bf k}\sigma},c^\dag_{5{\bf k}\sigma},
c^\dag_{6{\bf k}\sigma})$ is the six site basis, and
$\tilde{H}_{{\bf k}\sigma}$ is the Hamiltonian in $k$-space. The bulk energy bands as well as the bands on a strip geometry can be readily calculated. If the problem is solved on a two-dimensional strip, then periodic boundary conditions can be used in one direction (the direction parallel to the length of the strip). We diagonalize the Hamiltonian matrix using standard LAPACK routines.

\begin{figure}
\centering
\includegraphics[width=0.8\linewidth]{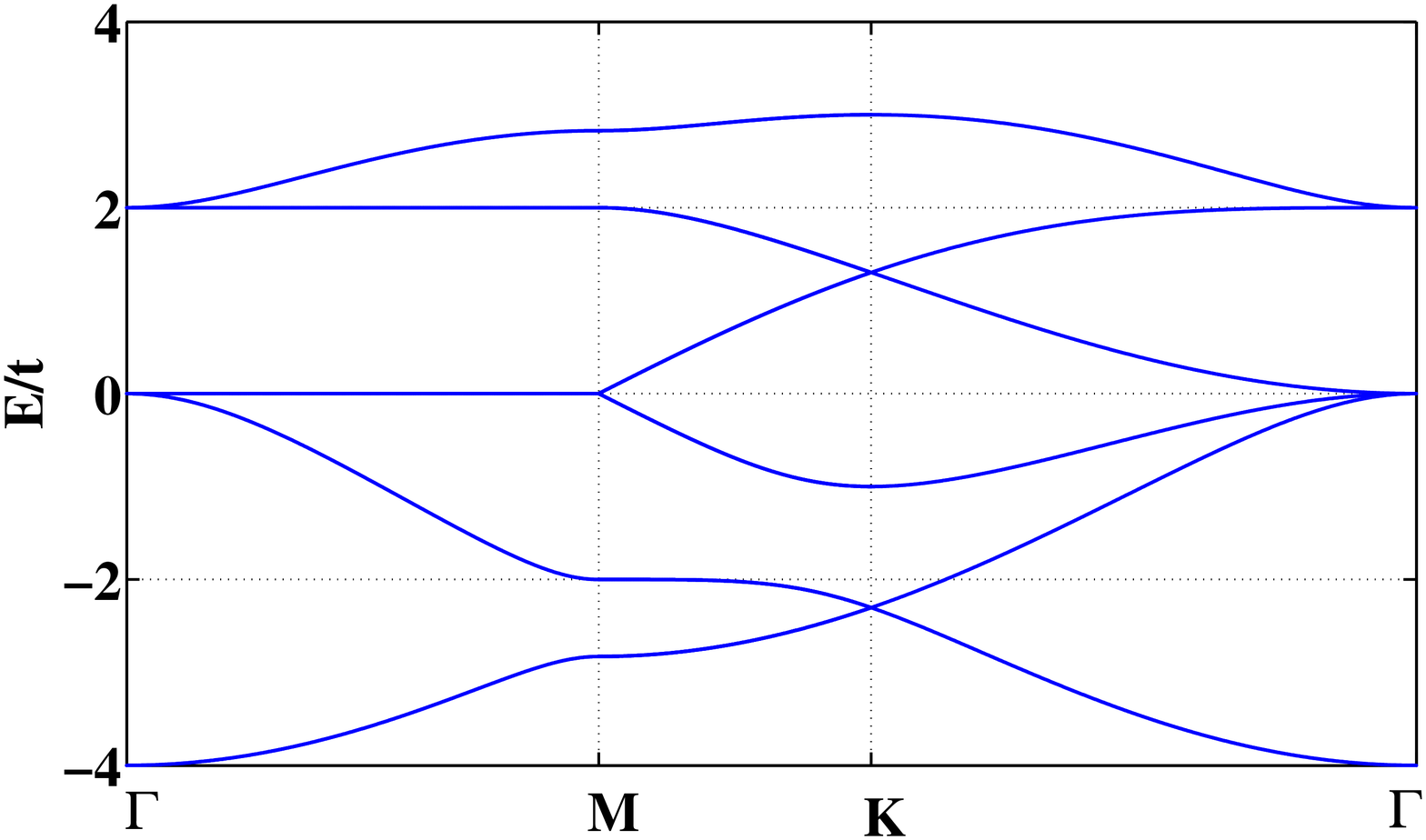}\\
(a)\\
\includegraphics[width=0.8\linewidth]{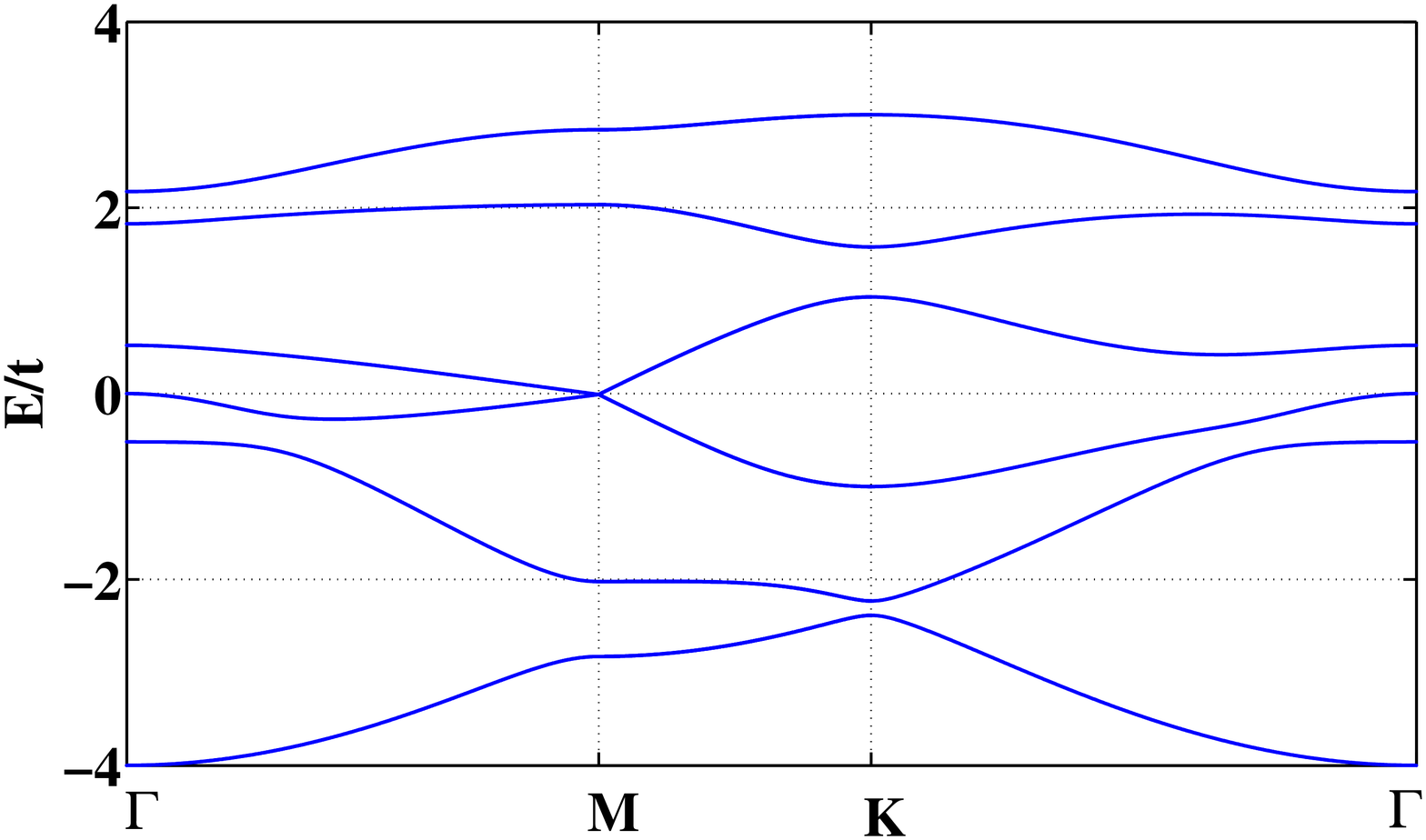}\\
(b)
\caption{The energy bands without any spin-orbit coupling and with finite spin-orbit coupling, given by Eq.\eqref{eq:H0}.  (a) Bulk
energy bands along high symmetry directions in the absence of spin-orbit coupling. Note that there is a Dirac point at $K$ for 1/6 and 2/3 filling, and a quadratic band touching point at $\Gamma$ for 1/2 and 5/6 filling. (The underlying lattice is triangular, as it is for the honeycomb lattice.) Note also the flat bands along the $\Gamma-M$ direction at 1/2 and 5/6 filling.  (b) The energy bands for finite spin-orbit coupling, with $t_2=t_3=0.1t$. It is clearly seen that the Dirac points at the K points and quadratic band touching point at the $\Gamma$ point are removed.}\label{fig:bands}
\end{figure}

\begin{figure}
\includegraphics[width=0.9\linewidth]{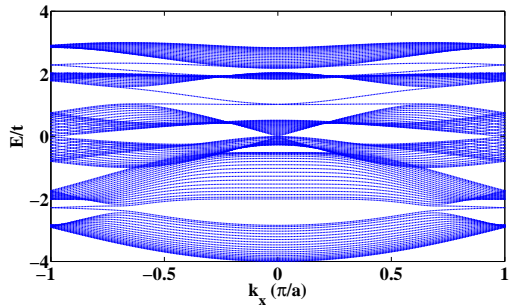}
\caption{The energy bands on a strip geometry for the case that $t_1=t$, $t_2=t_3=0.1t$.  At filling fractions 1/6 and 2/3 edge mode cross the band gap an odd number of times, which clearly reveals the topological insulator phases. Their identification is also confirmed with a direct evaluation of the $Z_2$ invariant.}\label{fig:bands2}
\end{figure}

The bulk energy bands and the energy bands on a strip geometry are shown in Figs.~\ref{fig:bands} and \ref{fig:bands2}, respectively, for different sets of parameters. Having 6 sites per unit cell, our tight binding model yields 6 bands whose are doubly degenerate due to spin degrees of freedom. Thus the filling factor is determined by counting how many of bands are fully occupied, which are 1/6, 1/3, 1/2, 2/3, 5/6.
If one imagines shrinking the triangular plaquettes on the ruby lattice down to a point, the links will look like those of the honeycomb, showing that the underlying Bravais lattice is triangular (since the underlying Bravais lattice of the honeycomb is triangular).  For the case $t_1=t$, $t_2=t_3=0$ shown in Fig.~\ref{fig:bands}(a) there is no spin-orbit coupling on the lattice, and at all filling fractions the model predicts metallic behavior. Also in Fig.~\ref{fig:bands}(a) we can see that when the energy is zero (1/2 filling), three bands are degenerate at the $\Gamma$ point.  For  $\frac 1 6$ filling and $\frac 2 3$ filling, there are Dirac points at the $K$ and $K'$ points. Spin-orbit coupling can gap out the spectrum at some fillings [see Fig.\ref{fig:bands}(b)] and drive the model into an insulator. The nature of these insulators can be further understood by looking at the spectrum of the system with edges as shown in Fig.~\ref{fig:bands2}. We see that at $\frac 1 6$ and $\frac 2 3$ filling,
there exist states at time reversal invariant points in $k$ space, that is, at the points where $k_x=0$ or $k_x=\pi/a$. These states traverse the bulk band gap and are composed of an odd number of Kramers pairs, indicating that the insulating phase is in fact topological.\cite{Kane:prl05,Kane_2:prl05}

A direct evaluation of the $Z_2$ invariant confirms this, as we will show in the next section. We found that the Dirac point at 2/3 filling in Fig.~\ref{fig:bands2} is a linear crossing point, but for the special case of $t_1=t$, $t_2=0$, and $-0.04t<t_3<0.04t$, the absolute value of the slope at $k_x=0$ is less than $10^{-4}$.  Therefore, in an experimental situation where temperature is finite (providing a low-energy cutt-off) it could be regarded as a quadratic crossing point. We also verified that a staggered potential with a magnitude as small as $0.01t$ which breaks the $C_4$ symmetry in the square plaquettes can enhance the slope by one order of magnitude. Our calculation is consistent with the discovery that extra symmetry of the underlying lattice may deform the shape of the Dirac node on the surface of 3D topological insulators giving rise to a warping effect\cite{Fu_hex:prl09} or a quadratic crossing in crystalline topological insulators.\cite{Fu:prl11} The flat crossing at 2/3 filling in our model may alter the low energy description of the edge modes and change their stability.\cite{Sun:prl09} In particular, this may make them more susceptible to magnetic ordering from Couloumb interactions.

Having established that our model supports the topological insulator phase,
we now turn to cases which explore the whole parameter space of the model and determine the phase diagrams at various filling fractions.

\section{Phase diagrams of the Model}
\label{sec:phase}
We are interested in determining the phases of the Hamiltonian \eqref{eq:H} as a function of filling fraction and the hopping parameters $t,t_1,t_2$, and $t_3$.  From earlier work\cite{Kane:prl05,Kane_2:prl05,Ruegg:prb10,Sun:prl09,Kargarian:prb10,Guo:prb09,Weeks:prb10} we know that we must consider three possible phases: (i) Conductor, (ii) Insulator, (iii) Topological insulator.

We determine the phase for a given set of parameters in the following way.  First we must determine whether the system is conducting or insulating. After choosing a filling fraction, we search for the bottom of the ``upper band" with respect to this filling fraction and the top of the ``lower band".  For example, at 1/2 filling the ``upper band" would be band 4 and the ``lower band" would be band 3, with the labeling starting at the lowest energy band and the counting increasing as one moves to higher energy bands. Finding the extrema of a given energy band is a formidable task since the band structure is rather complex and there are many local minima and maxima. We use an optimization algorithm called the differential evolution method.\cite{method} The method works well in most cases, but is not so efficient for some special parameters. In those cases, we use the software Mathematica to help us to determine the extreme values. If the system is insulating ({\em i.e.} there is a positive gap between the ``upper band" minimum and the ``lower band" maximum), then we calculate the $Z_2$ invariant according to the scheme proposed by Fu and Kane.\cite{Fu:prb07}

Taking the real-space triangular Bravais vectors of the ruby lattice as ${\bf a}_1=a\hat x$ and ${\bf a}_2=\frac{a}{2}\hat x +\frac{\sqrt 3 a}{2}\hat y$, the reciprocal lattice basis vectors are
\begin{eqnarray}
{\bf b}_1&=&\frac{2\pi}{a}\hat{x}-\frac{2\pi}{\sqrt{3}a}\hat{y},\nonumber\\
{\bf b}_2&=&\frac{4\pi}{\sqrt{3}a}\hat{y}.
\end{eqnarray}
 Since our model possess inversion symmetry, we will calculate the eigenvalues of the parity operator at the four time reversal invariant points in the $k$-space,\cite{Fu:prb07} that is,
\begin{equation}
{\bf b}=\frac{n_1}{2}{\bf b}_1+\frac{n_2}{2}{\bf b}_2,
\end{equation}
where
\begin{equation}
n_1, n_2=0, 1.
\end{equation}
From the eigenvalues of the parity operator at the time reversal invariant momenta, the $Z_2$ topological class can be determined as\cite{Fu:prb07}
\begin{equation}
(-1)^{\nu}=\prod_{a=1}^4 \delta_a,
\end{equation}
where
\begin{equation}
\delta_a=\prod_{m=1}^N \xi_{2m}(\Gamma_a).
\end{equation}
Here $\nu$ is the $Z_2$ topological invariant, $\Gamma_a$ is
one of the four time reversal invariant points defined as above
and $\xi_{2m}(\Gamma_a)$ is the eigenvalue of the parity operator
for the Bloch wave function of the $2m$-th occupied band
at the time reversal invariant point $\Gamma_a$.

Because of the large parameter space of the model, we must choose different ``cuts" of the parameters to explore the phase diagrams.  We will use the methods described above to determine the phases at filling fractions for different hopping values.

\subsection{Phase diagrams for $t_2=t_3$}

\begin{figure}[h]
\includegraphics[width=0.45\linewidth]{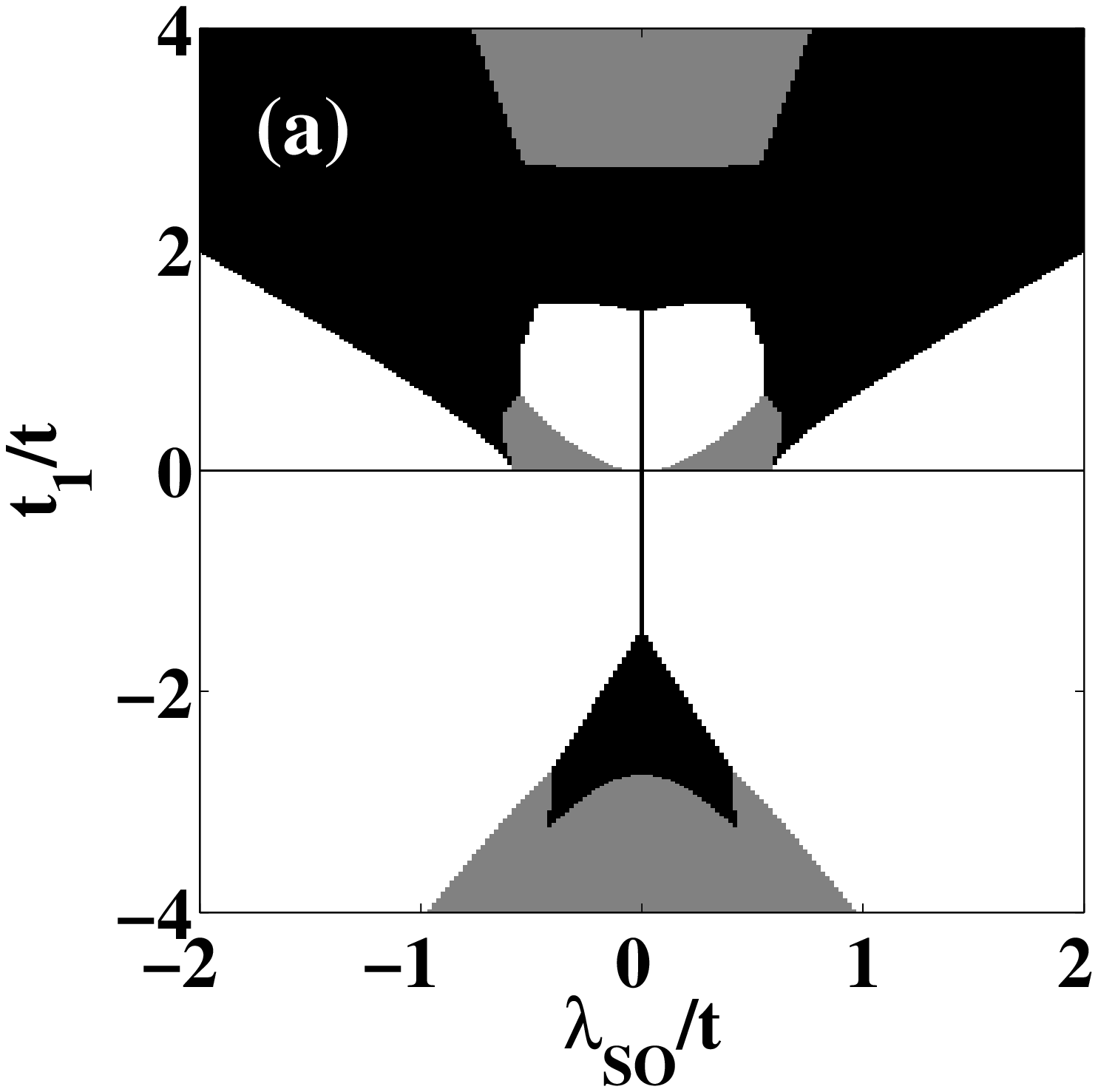}
\includegraphics[width=0.45\linewidth]{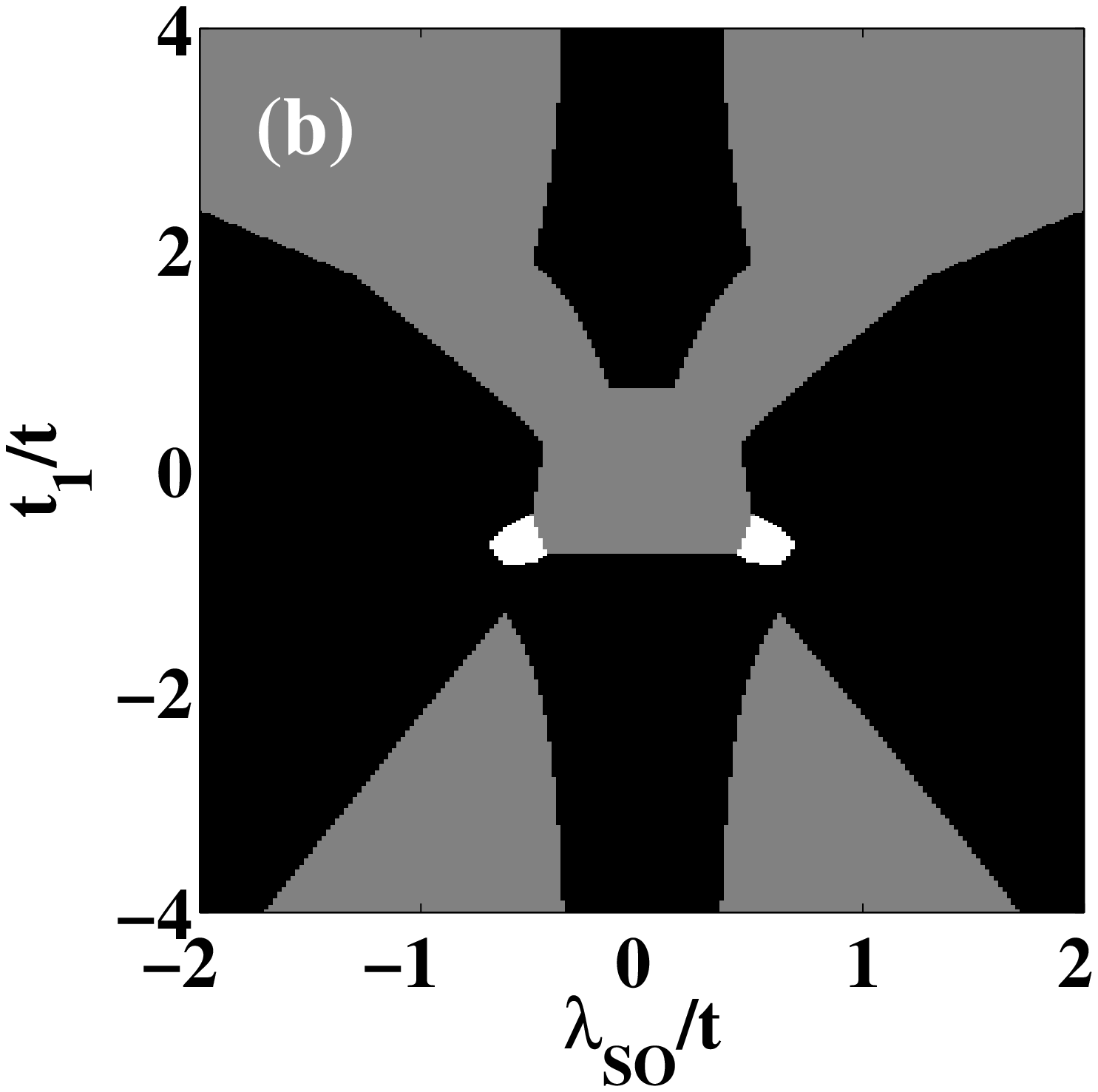}\\
\includegraphics[width=0.45\linewidth]{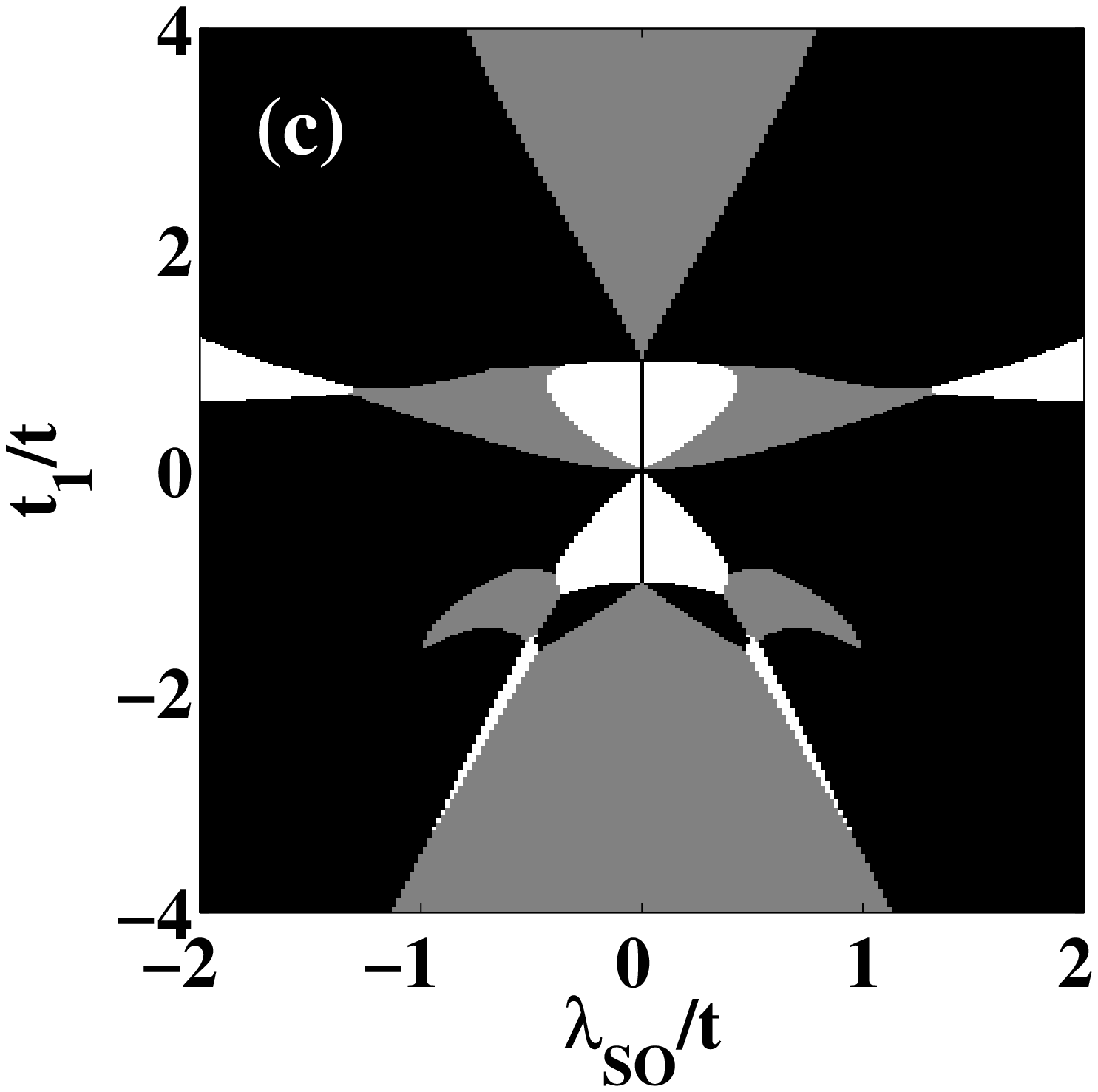}
\includegraphics[width=0.45\linewidth]{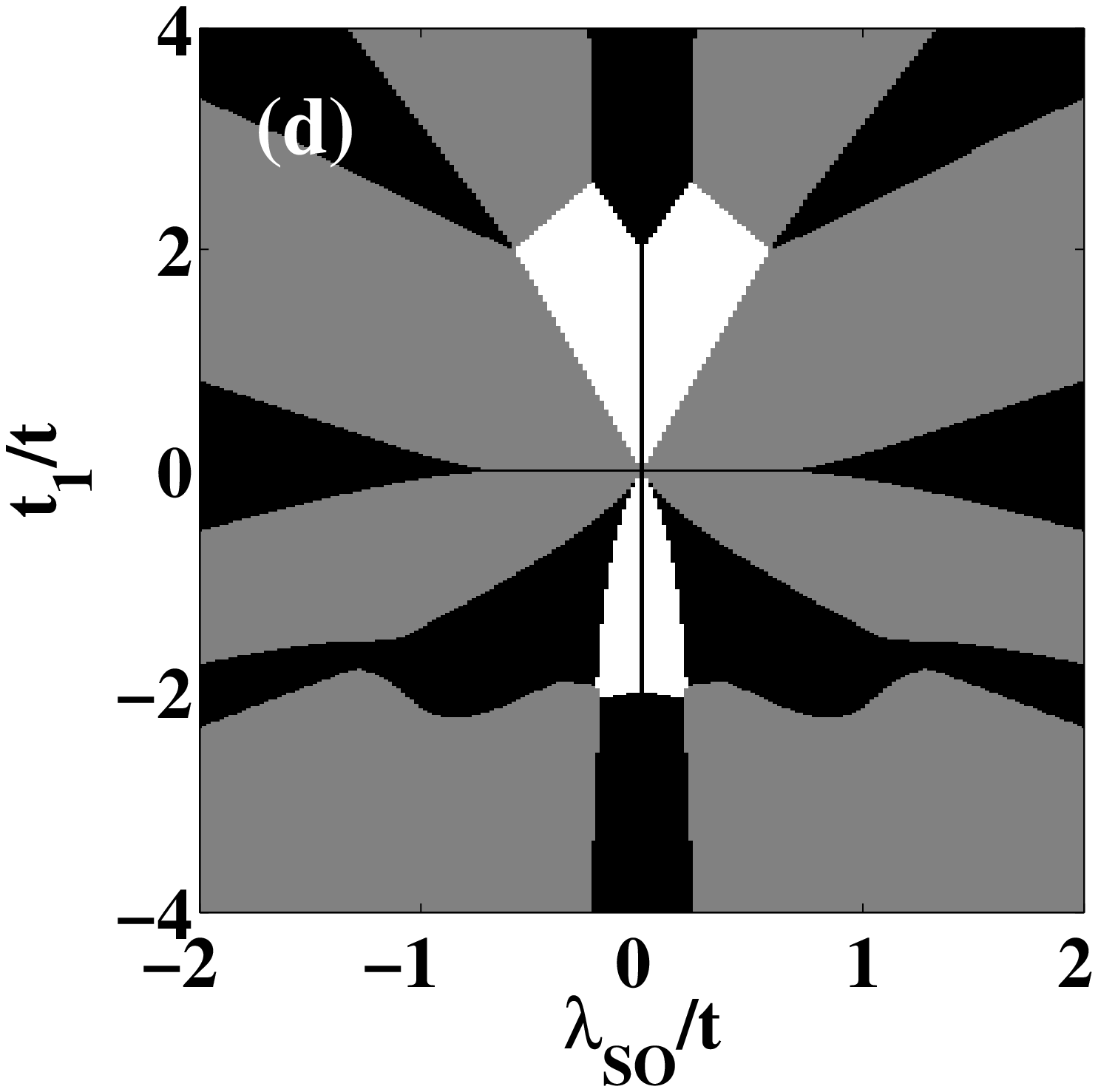}\\
\includegraphics[width=0.45\linewidth]{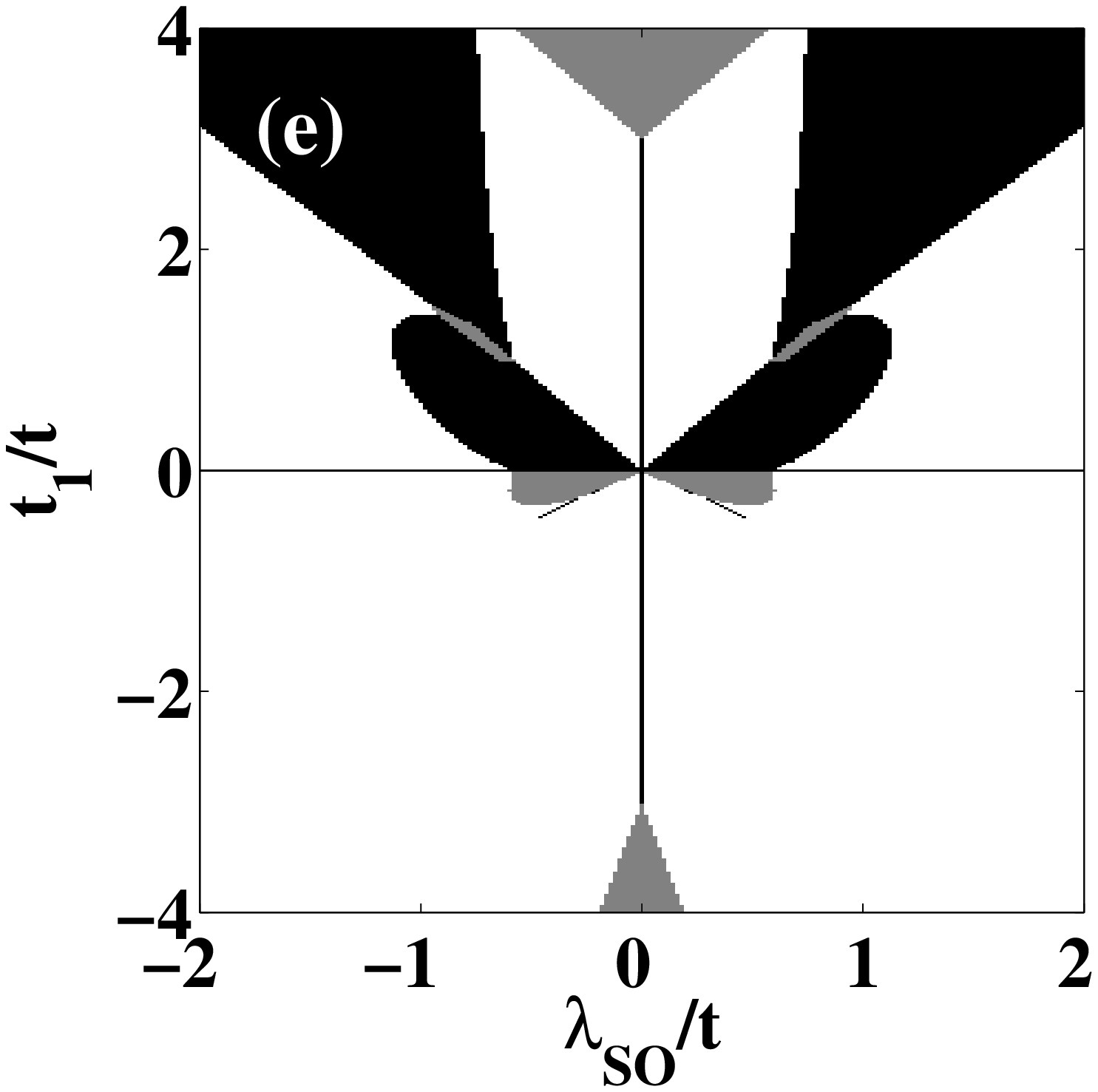}
\caption{Phase diagrams for different filling fractions. The figures (a)-(e) are for the
filling fraction from 1/6 to 5/6, with filling fraction increasing in units of 1/6. For each filling, the vertical and horizontal axes measure $t_1$ and $t_2=t_3=\lambda_{\rm SO}$, in units of $t$, respectively. Color coding is: Black=Conductor, Grey=Insulator, White=Topological Insulator. }\label{fig:t2t3}
\end{figure}

We begin by fixing the second neighbor hopping values $t_2=t_3=\lambda_{\rm SO}$ in \eqref{eq:H_SO}. Thus we take uniform spin orbit coupling $\lambda_{\rm SO}$ and inter-triangle hopping $t_1$ as tuning parameters which are expected to drive the system into different phases. Fig.~\ref{fig:t2t3} depicts the phase diagrams for filling fractions 1/6, 1/3,1/2,2/3, and 5/6.  {\em In the figure, and all related figures that follow in the paper the color coding is: Black=Conductor, Grey=Insulator, White=Topological Insulator. }

A variety of phases is seen at different fillings. Starting with $\frac 1 6$ filling, we see that the
middle axis $\lambda_{\rm SO}=0$ is never in the topological insulator phase. This is easy to understand because in the absence of spin-orbit coupling, the system can only be a metal
or trivial insulator. There also exists a horizontal straight line on the phase diagram which corresponds to the metallic phase when the inter-triangle hopping is zero. It is noteworthy that large areas of the phase diagram are occupied by the topological insulator. That is consistent with the observation of Dirac points at $\frac 1 6$ filling for $t_2=t_3=0$ in the bulk energy bands and the
crossing edge modes on the strip geometry for $\lambda_{\rm SO}\neq0$.  The phenomenology is similar to the case of the decorated honeycomb lattice model.\cite{Ruegg:prb10}

At $\frac 1 3$ filling most areas are either a conductor or a trivial insulator,
which is consistent with the lack of Dirac cones or quadratically touching points
in the bulk energy bands when there is no spin-orbit coupling. However, as the spin-orbit coupling is turned on, at some special parameters the system can still be a topological insulator which are shown as white spots in Fig.~\ref{fig:t2t3}(b).

At $\frac 1 2$ filling the phase diagram appears rather complex. In most
areas it is a conductor, while for some areas the trivial insulator
and the topological insulator can be distinguished.
The topological insulator phase also appears on some narrow areas close
to the region around $\lambda_{\rm SO}=\pm0.7t$ and $t_1=-2.0t$.

The phase diagram for $\frac 2 3$ filling consists mainly of the trivial insulator and the
metallic phase. Nevertheless small regions around the center of the phase diagram occurring at small values of spin-orbit coupling present the topological insulator phase.  For much of the phase diagram, the conducting and trivial insulating phases
occur alternately when the hopping strength and/or spin-orbit coupling
are increased. The crossing states at $k=0$ in the calculation on the strip geometry shown in Fig.~\ref{fig:bands2}, lies in the narrow region of the topological insulating phase.

Finally, for $\frac 5 6$ filling most of the area is occupied by the topological insulator, in strong contrast to the former cases. This is also clearly seen from the calculation in the strip geometry shown in Fig.~\ref{fig:bands2} combined with the less stable quadratic band touching points\cite{Sun:prl09,FZhang:prl11} (compared to Dirac points) that are clearly seen at 5/6 filling in Fig.~\ref{fig:bands}(a) at the $\Gamma$ point.

\subsection{Phase diagrams for fixed $t_1$}

As we emphasized earlier, the model \eqref{eq:H_SO} supports two types of spin-orbit couplings, $t_2,t_3$. In this section we investigate the interplay between theses couplings and the resulting phases.  We describe phase diagrams with the same color code as before: Black=Conductor, Grey=Insulator, White=Topological Insulator.
We assume that the hopping between triangles, $t_1$, is set to a fixed value, and
the spin-orbit coupling $t_2$ and $t_3$ are independently varied. This type of phase diagram
takes a slice in the three dimensional parameter space $(t_1/t,t_2/t,t_3/t)$. Such phase diagrams reveal even more features than those presented earlier with $t_2=t_3$ (Fig.\ref{fig:t2t3}). In particular, we find that the two second-neighbor spin-orbit coupling terms can ``compete" with each other and drive the system out of the topological insulator phase, even though each individually would place it there.

\begin{figure}
\includegraphics[width=0.45\linewidth]{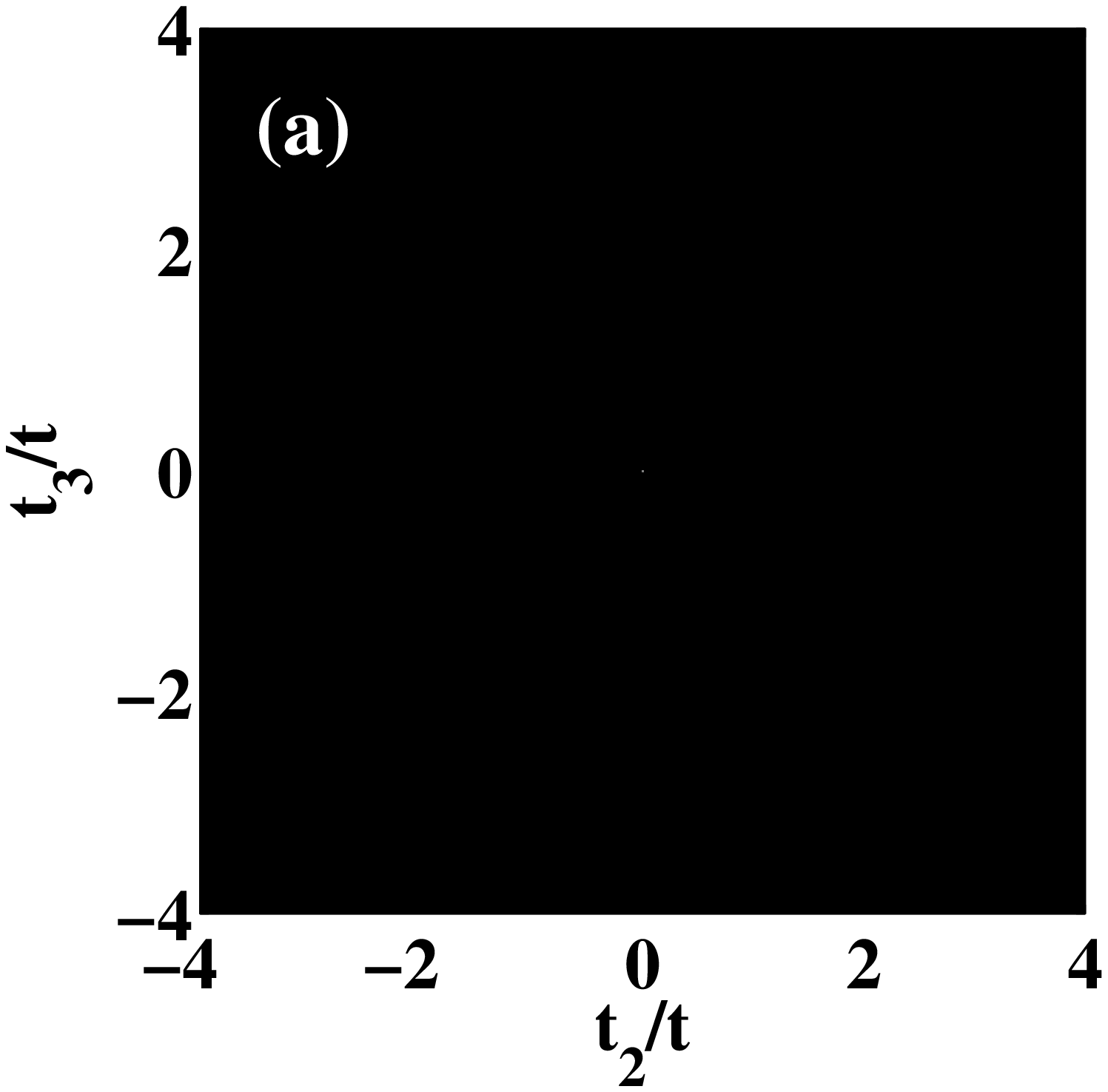}
\includegraphics[width=0.45\linewidth]{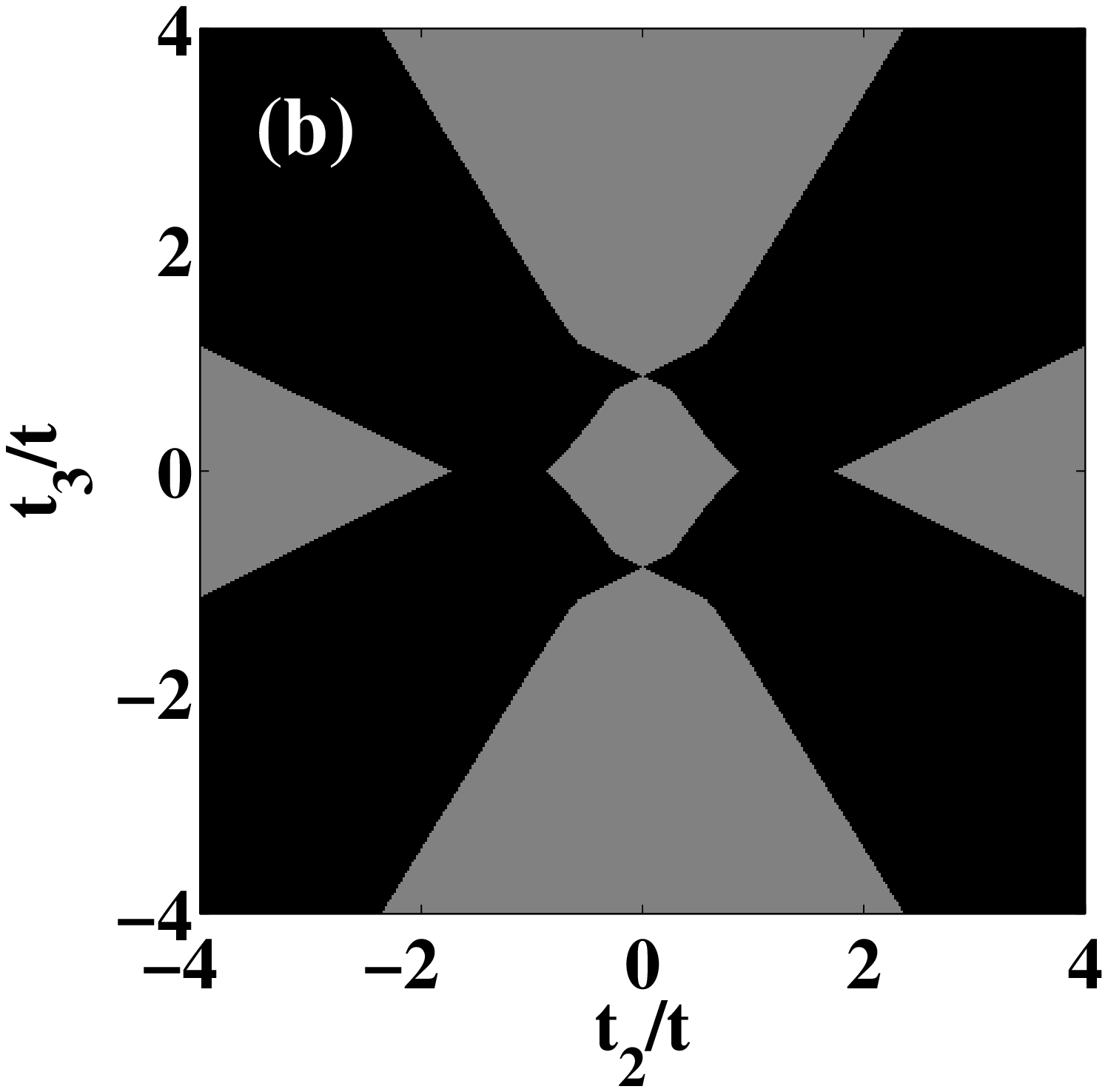}\\
\includegraphics[width=0.45\linewidth]{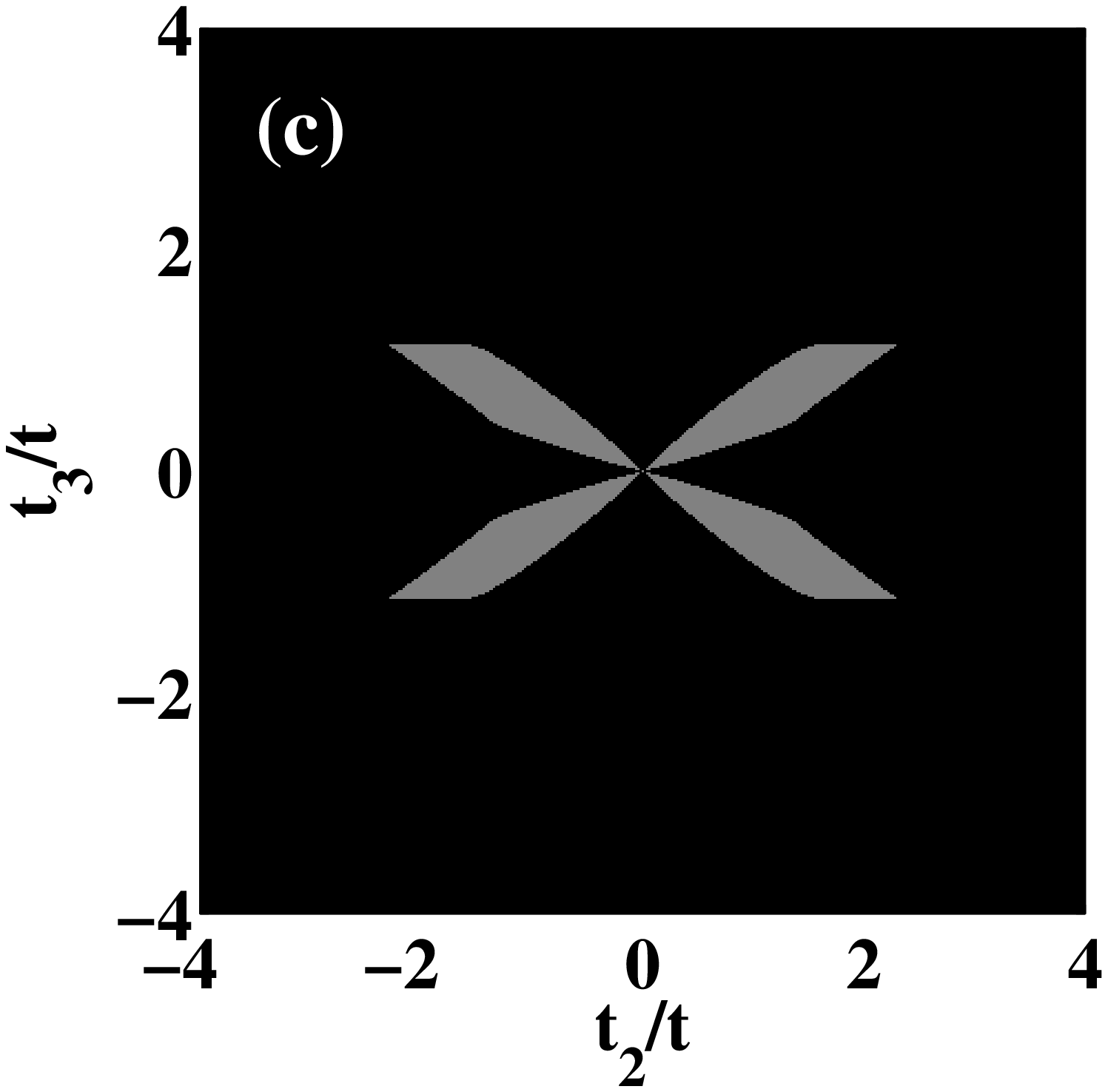}
\includegraphics[width=0.45\linewidth]{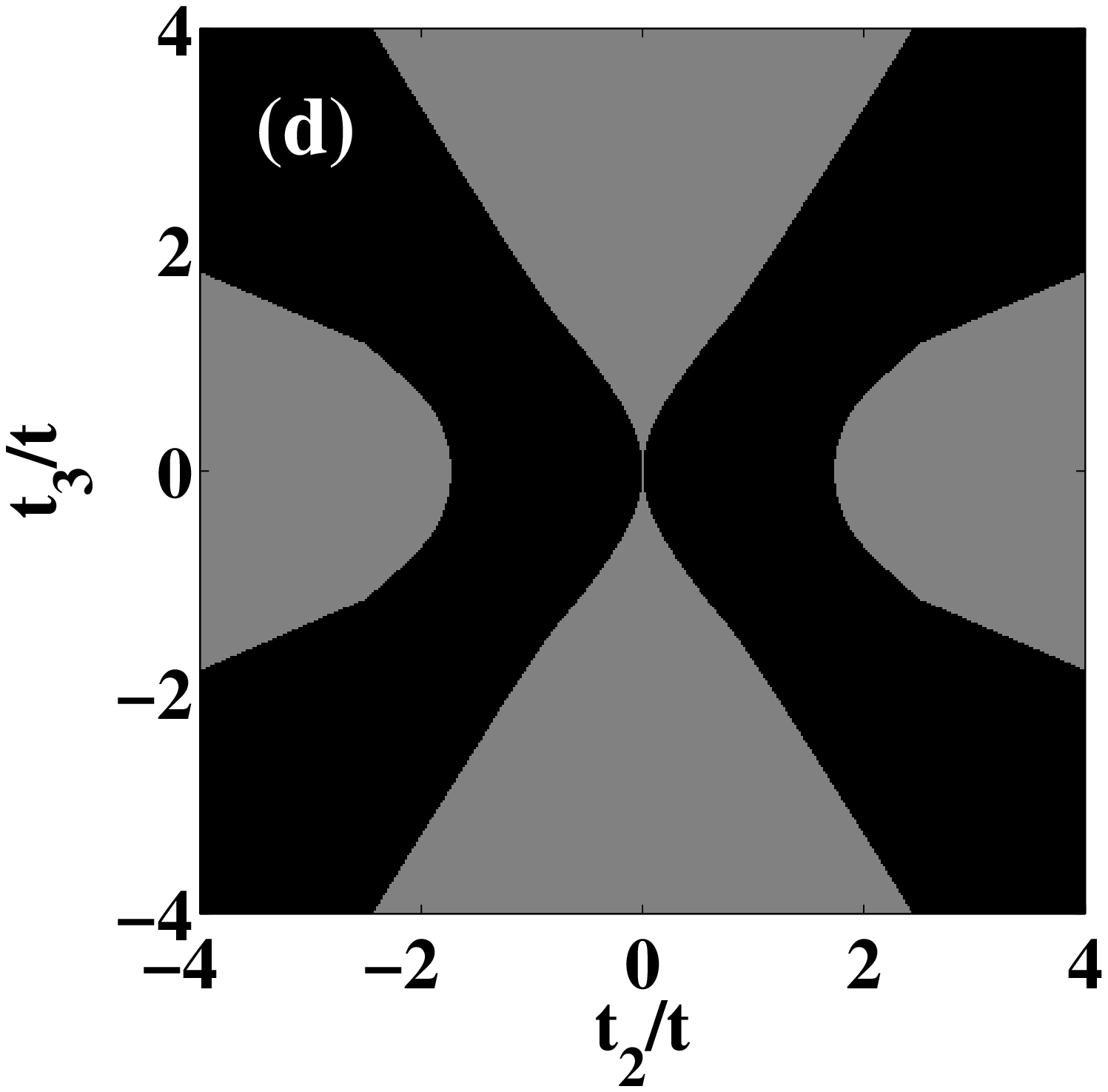}\\
\includegraphics[width=0.45\linewidth]{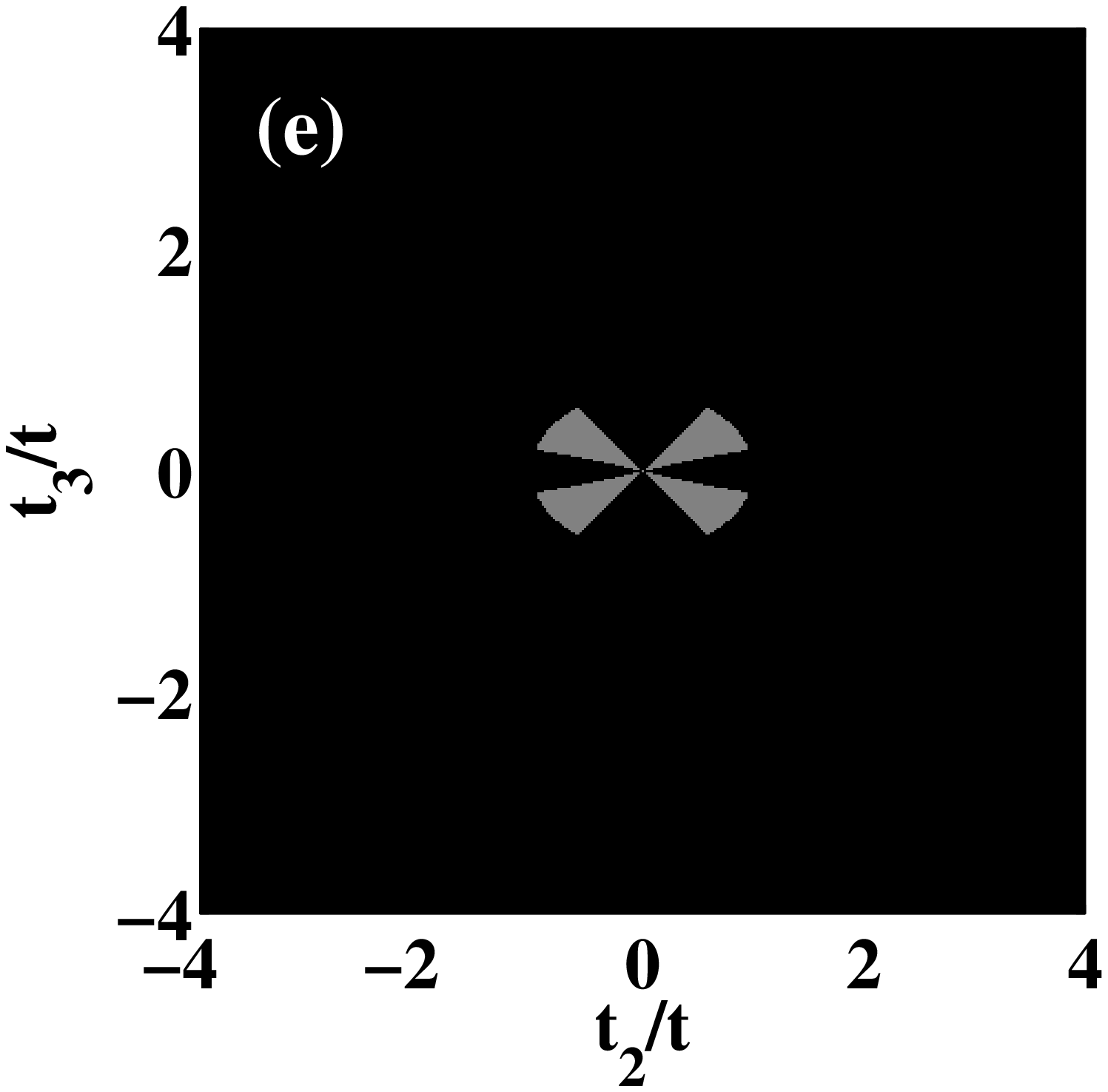}
\caption{Phase diagrams for $t_1=0$. The horizontal axis is $t_2/t$, and
the vertical axis is $t_3/t$. The figures (a)-(e) are for the
filling fraction from 1/6 to 5/6, increasing in units of 1/6. For all fillings the origin $t_2=t_3=0$ is a trivial insulator.}\label{fig:0t}
\end{figure}

\begin{figure}
\includegraphics[width=0.45\linewidth]{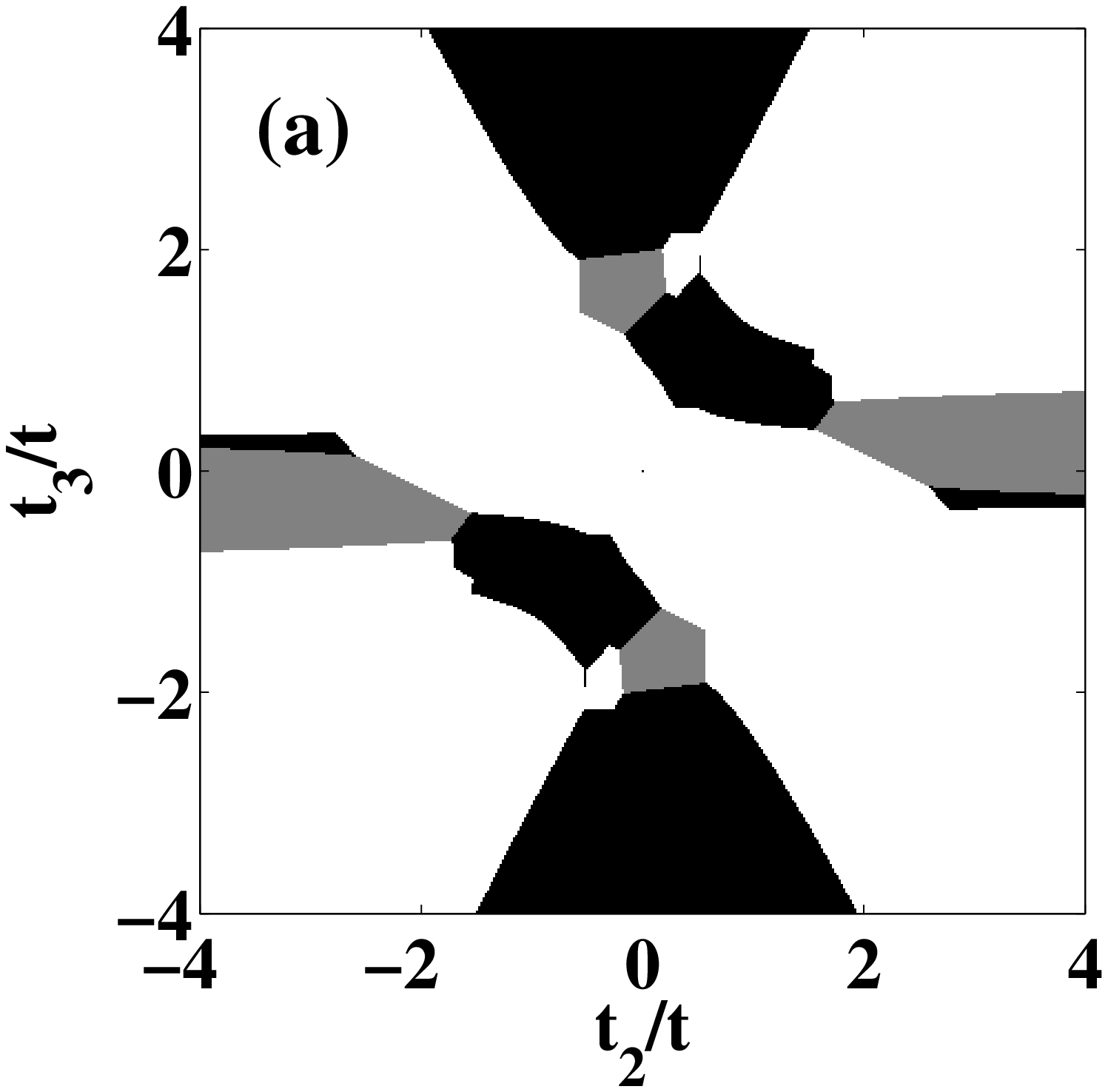}
\includegraphics[width=0.45\linewidth]{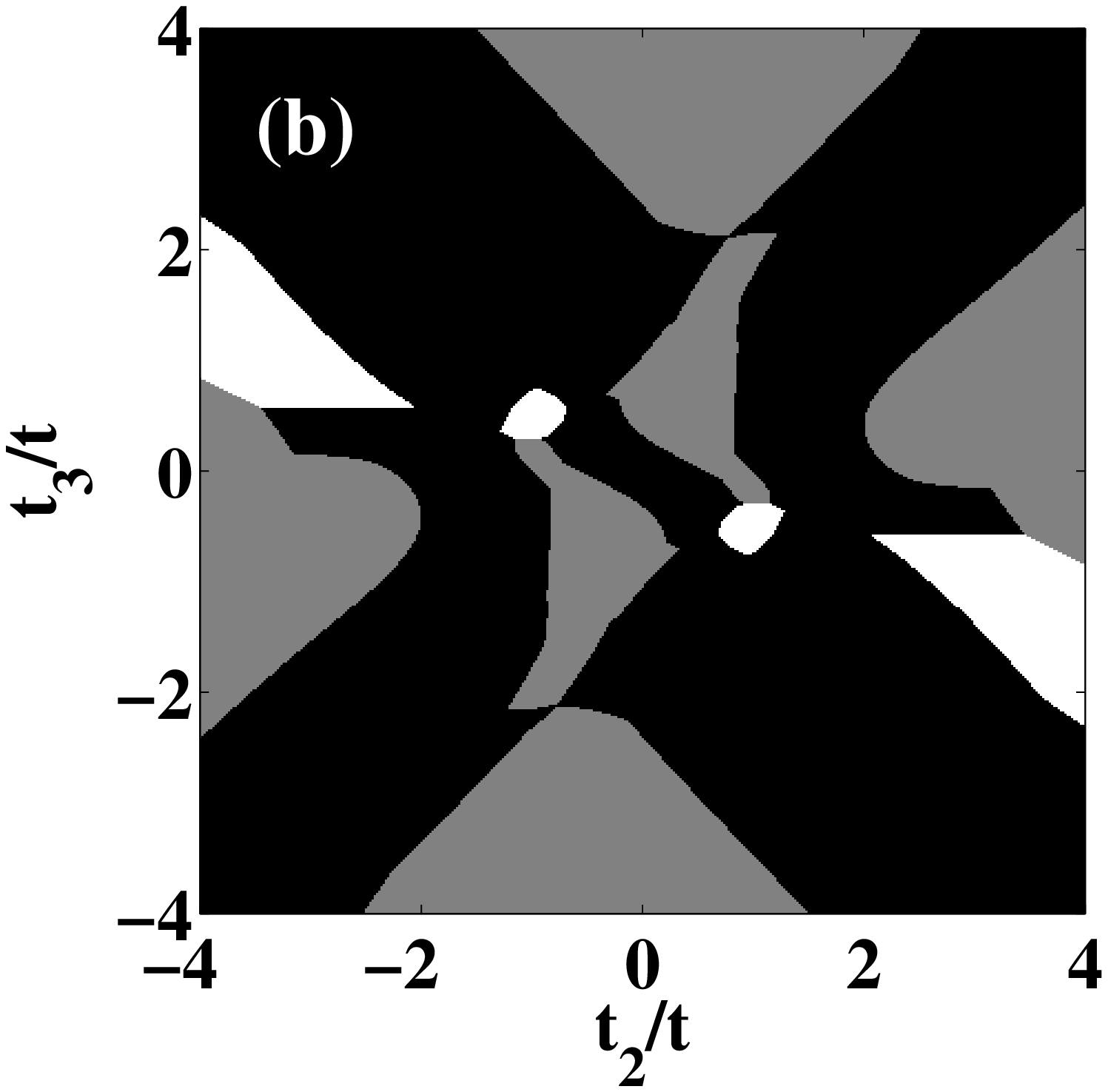}\\
\includegraphics[width=0.45\linewidth]{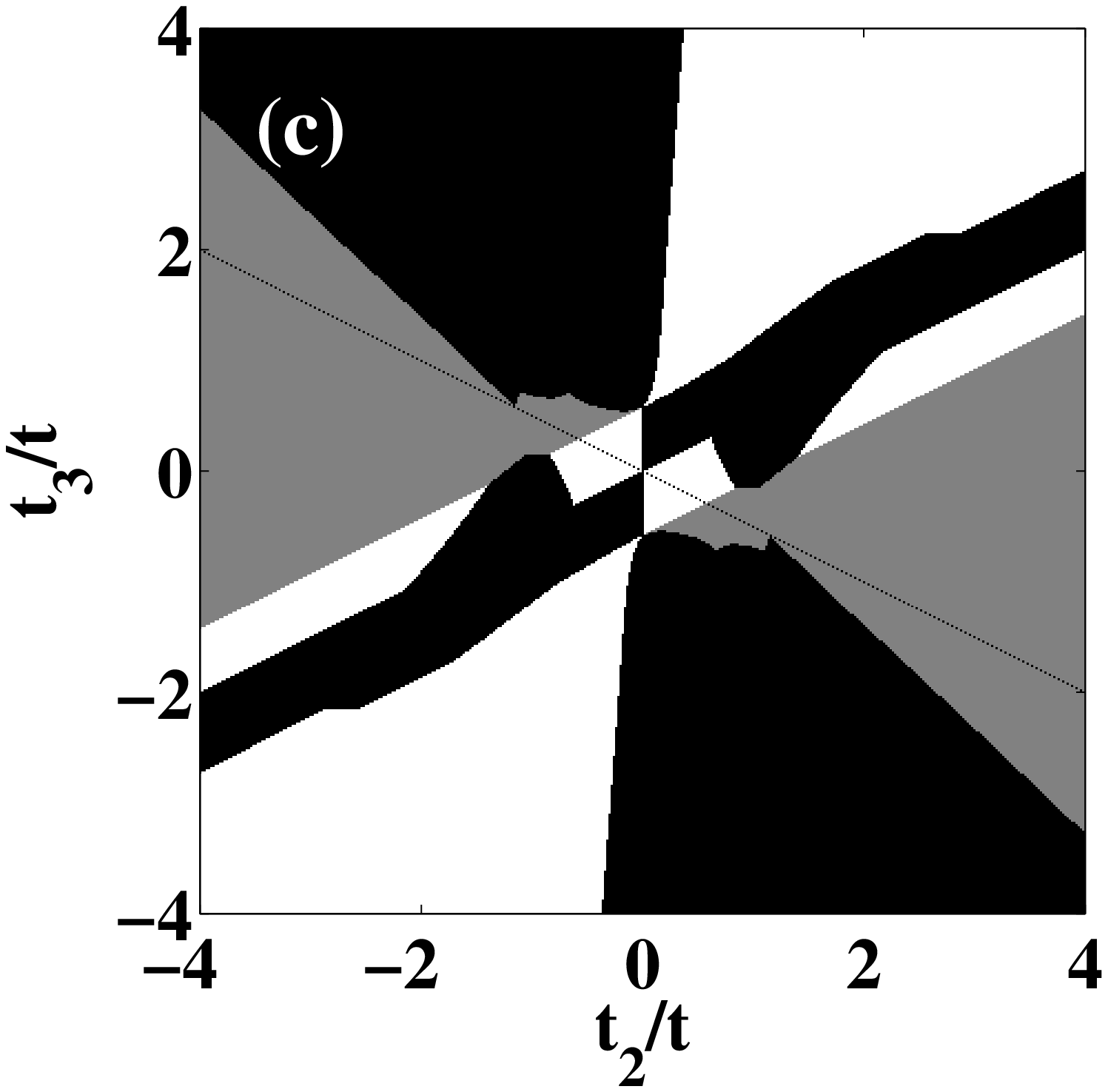}
\includegraphics[width=0.45\linewidth]{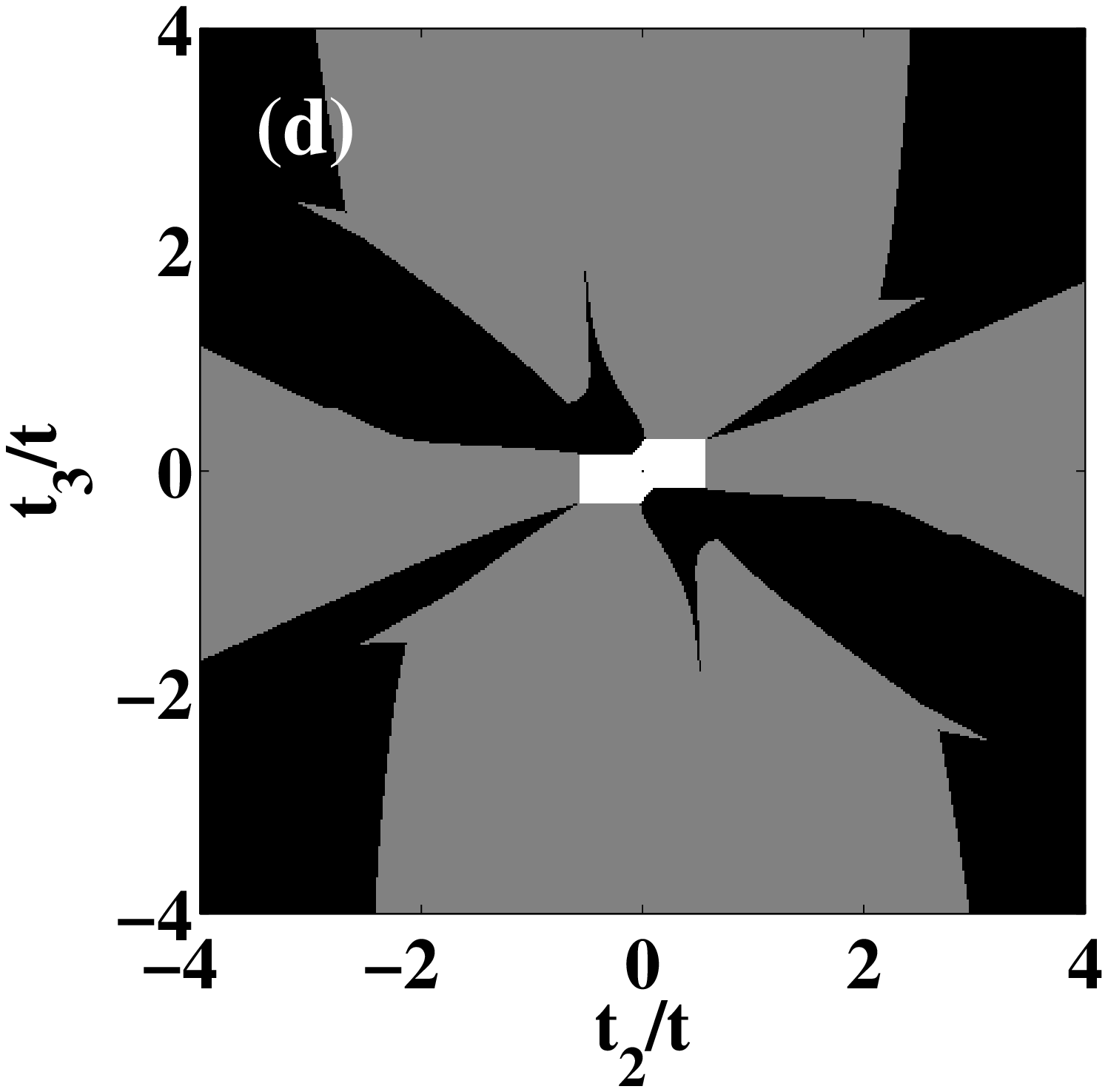}\\
\includegraphics[width=0.45\linewidth]{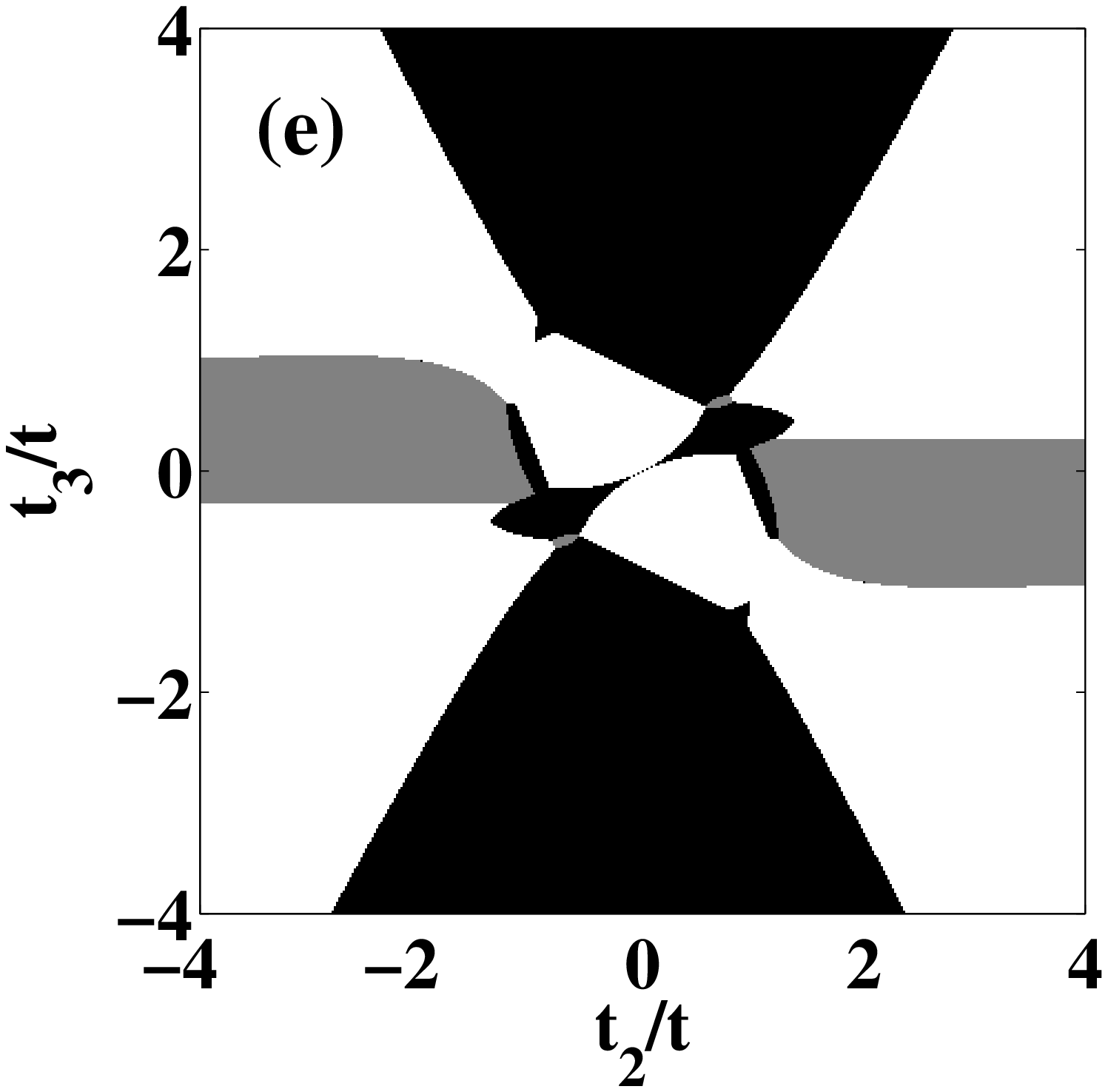}
\caption{Phase diagrams for $t_1=t$. The horizontal axis is $t_2/t$, and
the vertical axis is $t_3/t$. The figures (a)-(e) are for the
filling fraction from 1/6 to 5/6, increasing in units of 1/6.}\label{fig:1t}
\end{figure}

We consider two cases: $t_1=0$ and $t_1=t$. The corresponding phase diagrams are shown in Figs.~\ref{fig:0t}-\ref{fig:1t}. The phase diagrams illustrated in Fig.~\ref{fig:0t} in which $t_1=0$ are a special case. First note that in the absence of second-neighbor hopping it is clear the model must be in a trivial insulator phase since the only hopping is around triangular plaquettes in the lattice.  Thus, the origin at all filling fractions is a trivial insulator.  However, away from the origin the behavior is rather different at different filling fractions.  For example, close to the origin the model is in a trivial insulating state at 1/3 filling, while it remains a metal nearby at 2/3 filling (for the most part). Interesting, at $\frac 1 6$ filling only the metallic phase appears. Also, at 1/2 and 5/6 filling the phase diagram is mostly metallic.

In Fig.~\ref{fig:1t} we show the phase diagrams for $t_1=t$. Although those phase diagrams show a complex evolution with the hopping parameters, we can still draw some clear conclusions from them. For example, filling fraction 1/6 and 5/6 are rich in regions of topological insulator, while filling fractions 1/3 and 2/3 are poor in regions of topological insulator.  Filling fraction 1/2 tends to have roughly one-third to half of the parameter space occupied by the topological insulator phase. Ultimately, the explanation for these behaviors comes from the band structure in the presence of second-neighbor hopping.  However, as shown in Fig.~\ref{fig:bands}(a), the bands at filling fraction 1/6 and 2/3 both have a Dirac cone at the $K$-point so the naive expectation might be for these two fillings to respond similarly when second-neighbor spin-orbit couplings are added.  Our calculations clearly indicate this is not the case and the actual evolution is rather complicated. This feature should be borne in mind when viewing the present model as the result of a self-consistent Hamiltonian in which spin-orbit coupling was spontaneously generated from interactions.\cite{Raghu:prl08,Sun:prl09,Wen:prb10,Zhang:prb09,Liu:prb10} We note that if the signs of any two among $t_1$, $t_2$, and $t_3$ are reversed, the figures show the phase of the system is unchanged.  For example, if one flips the phase diagram for $t_1=t$ (See Fig.~\ref{fig:1t}) around the axis $t_2=0$ or $t_3=0$, the phase diagram obtained is identical to the phase diagram for the case $t_1=-t$. This symmetry is easy to verify from the Hamiltonian \eqref{eq:H}. Indeed, the sign of $t_1$ can be absorbed via a simple gauge transformation of electron operators. The latter is defined as follows: $c^{\dag}(c)\rightarrow -c^{\dag}(-c)$ for up triangles and $c^{\dag}(c)\rightarrow c^{\dag}(c)$ for down triangles. This transformation flips the signs of $t_1$ and $t_3$. Thus this simple argument implies that the phase diagram is symmetric via such transformations. Moreover, if the signs of $t_2$ and $t_3$ are flipped, the phase diagram will not change, which can be simply understood by noting that this flipping can also be absorbed into the magnetic fields for different spin species, i.e. $\nu_{ij} \rightarrow -\nu_{ij}$.

We also studied other cases with different values of inter-triangle hopping. For example, when $t_1=2t$ and
the filling fraction is $\frac 5 6$, if $t_2=t$ and $t_3=0$, or $t_2=0$ and $t_3=0.4t$,
the system is topological insulating, while if $t_2=t$ and $t_3=0.4t$, the
system is a metal. A possible explanation for such phenomena is the inner
magnetic field created by the two different types of spin-orbit coupling may
have opposite contribution to the phase of the hopping terms, and therefore may cancel each other for some parameter values.  A similar inverse case also exists where a single type of spin-orbit coupling does not result in the topological insulator phase while if both types of spin-orbit coupling are present, the system becomes a topological insulator.

\begin{figure}[h]
\includegraphics[width=0.45\linewidth]{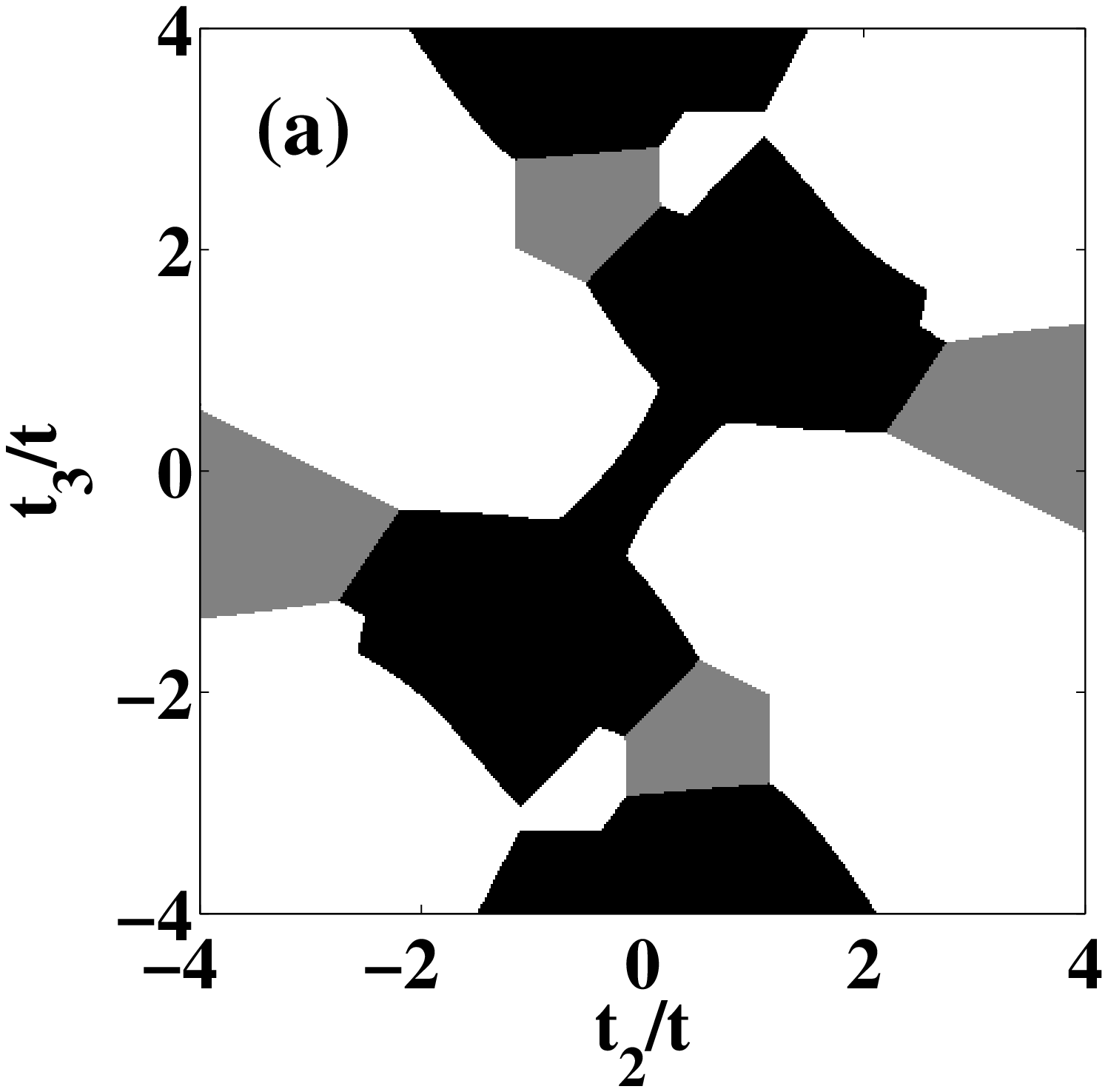}
\includegraphics[width=0.45\linewidth]{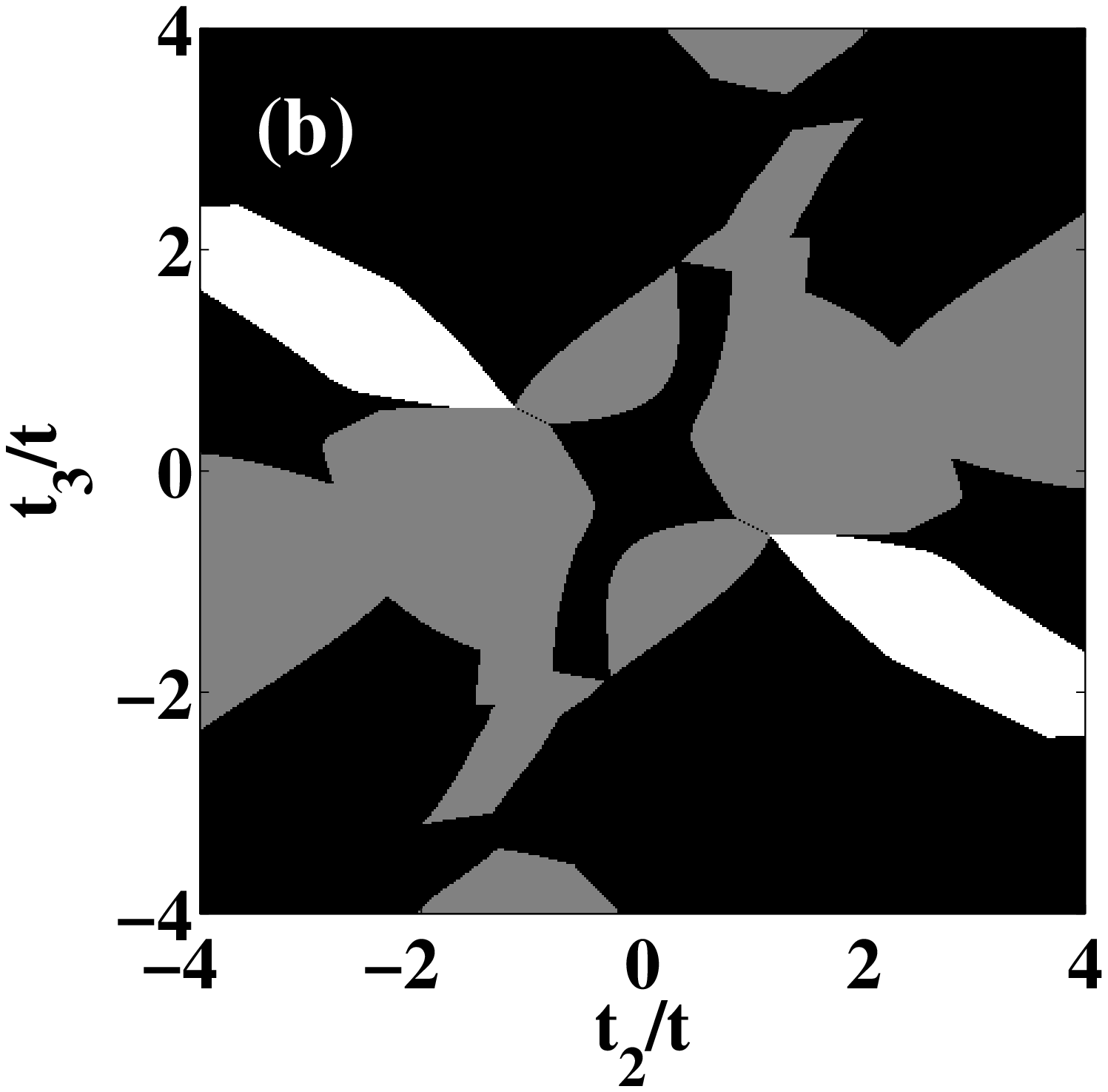}\\
\includegraphics[width=0.45\linewidth]{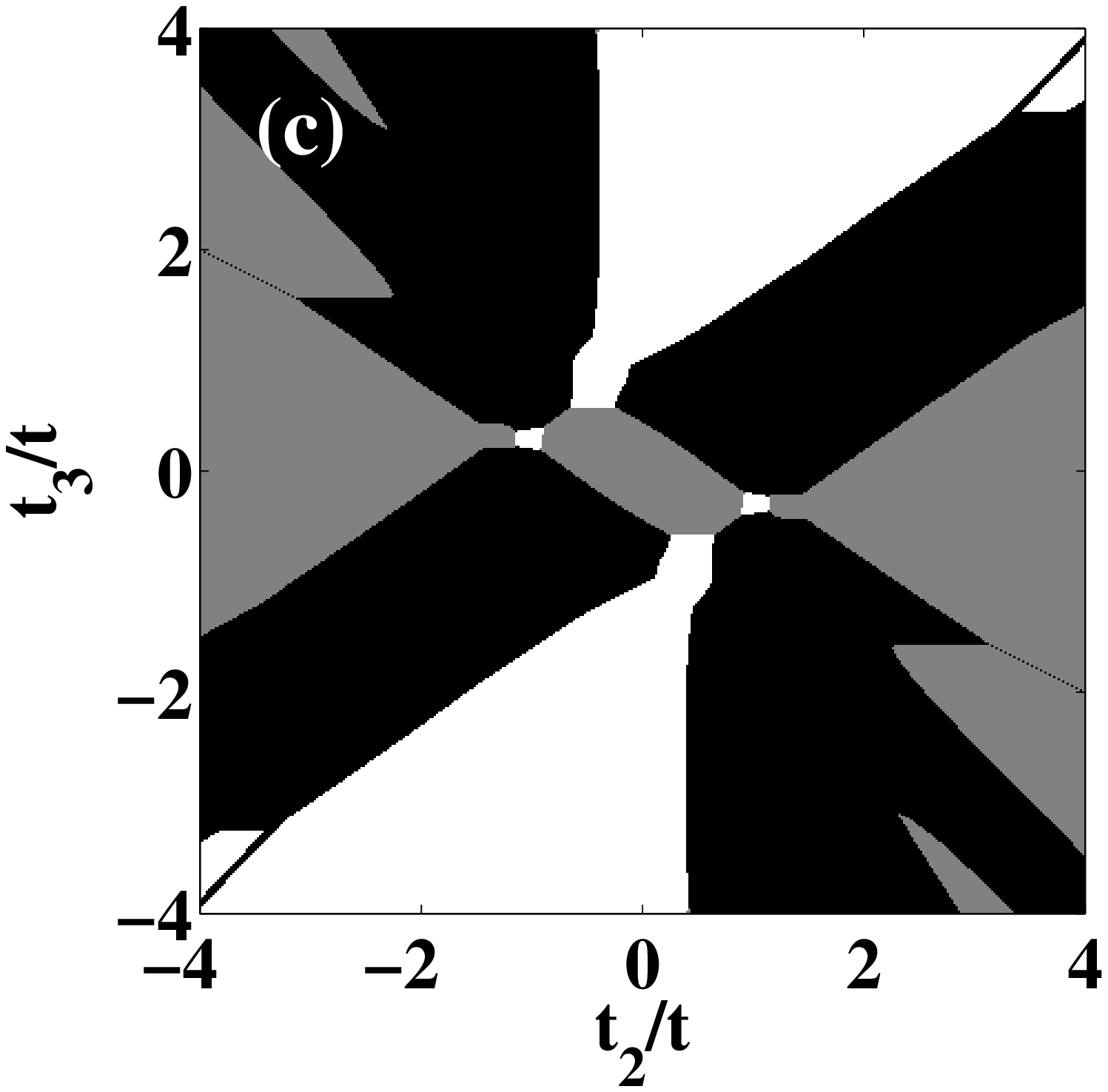}
\includegraphics[width=0.45\linewidth]{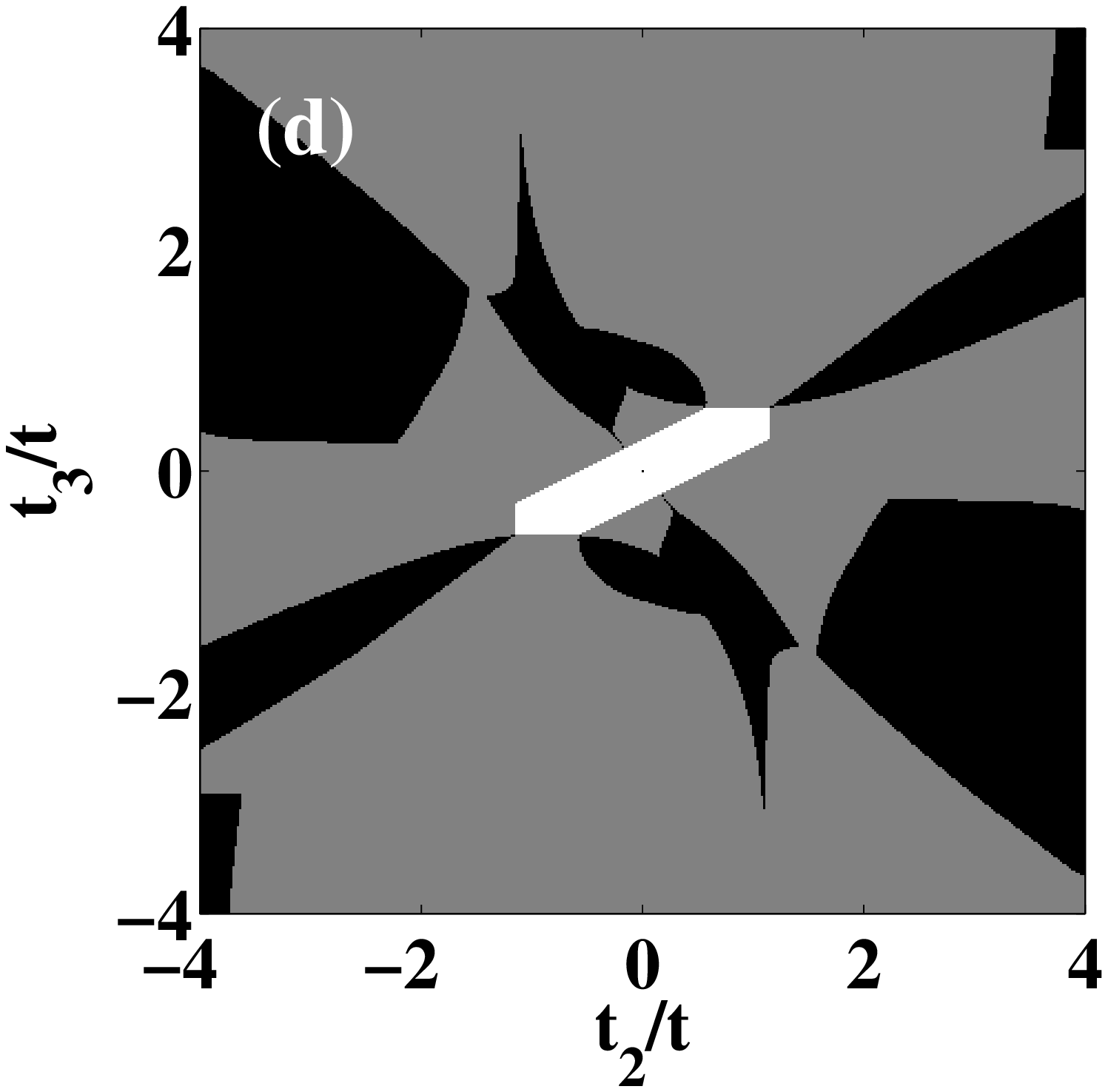}\\
\includegraphics[width=0.45\linewidth]{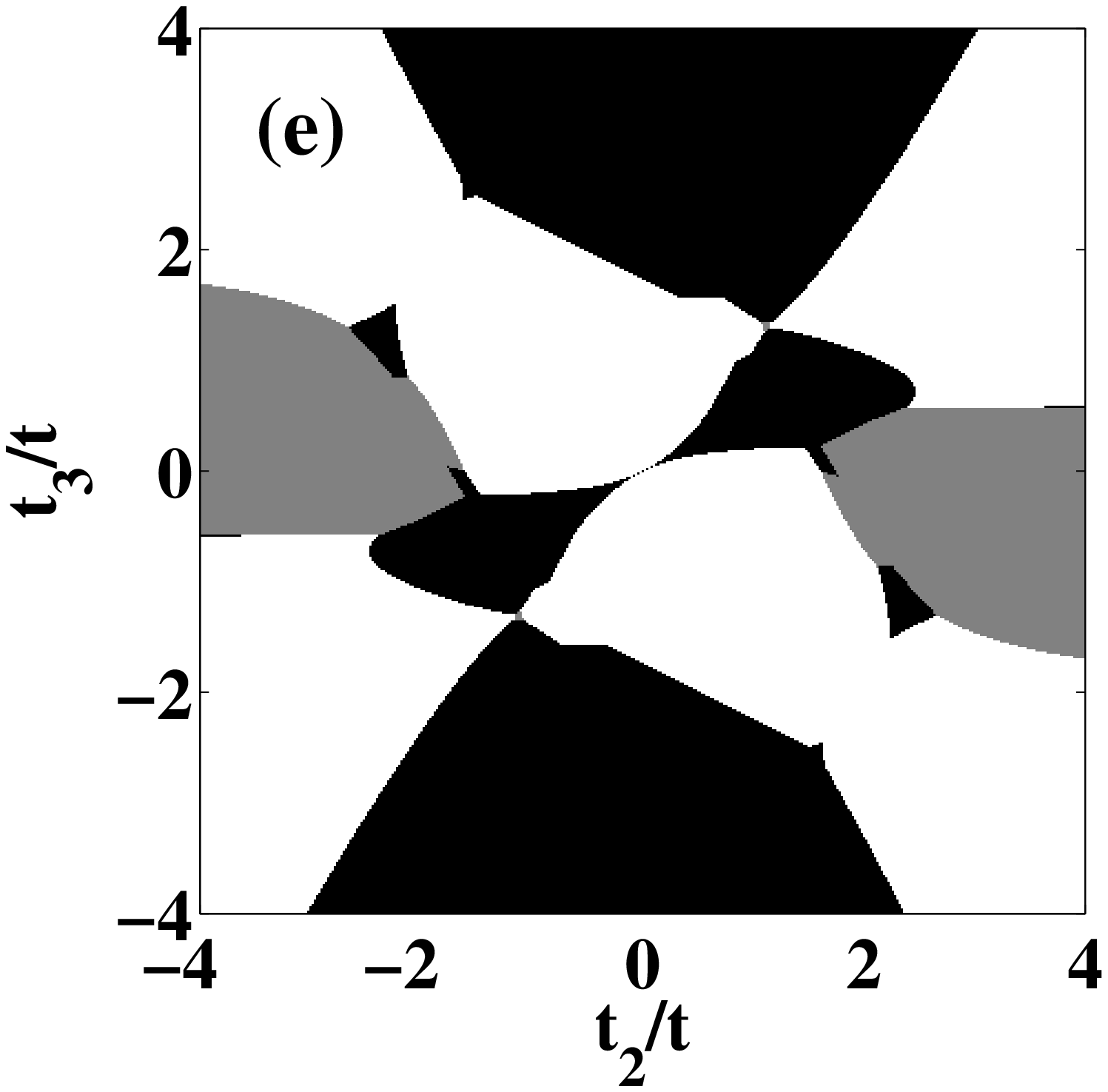}
\caption{Phase diagrams for $t_1=2t$. The horizontal axis is $t_2/t$, and
the vertical axis is $t_3/t$. The figures (a)-(e) are for the
filling fraction from 1/6 to 5/6, increasing in units of 1/6.}\label{fig:2t}
\end{figure}

\section{Flat band fractional quantum Hall effect}
\label{sec:FQHE}

In recent years, lattice models with flat bands have attracted attention for a number of reasons, among them are enhanced interaction effects.\cite{Ohgushi:prb00,Bergman:prb08,Xiao:prb03,Wu:prl07,Green:prb10,Kapit:prl10}  Since the scale of the kinetic energy is set by the band width, if the band width vanishes, any residual inter-particle interactions will be ``large" and may drive the system into a strongly correlated state such as a Wigner crystal\cite{Wu:prl07} or a fractional quantum Hall state.\cite{Kapit:prl10}

Flat band lattice models in which the flat bands possess a finite Chern number are thought to be excellent candidates for systems that might realize a fractional quantum Hall effect when the flat band is partially filled.\cite{Sun10,Neupert11,Tang10,Wang11}  Indeed, exact diagonalization studies on the checkerboard lattice\cite{Sheng11, Wang11} and other lattices\cite{Neupert11, Wang11} support the development of a fractional quantum Hall state at certain filling fractions.

In order to maximize the effectiveness of the interactions in driving a fractional quantum Hall state, the flat band should be as flat as possible {\em and} the band gap to the next band should be as large as possible.  (Virtual transitions to higher lying bands with finite Chern numbers are more effective at disrupting the fractional quantum Hall state than those with zero Chern number.)  If we call the bandwidth of the flat band $W$ and the energy difference between the lowest point of the band above the flat band and the highest point of the flat band $E_g$, the figure of merit is $E_g/W$, and this should be much larger than unity.  If the characteristic strength of the inter-particle interactions is $U$, the regime $W \ll U \ll E_g$ will involve little mixing from the higher bands and the physics will be dominated by the interactions in the flat band with a finite Chern number.

In this section, we show that the ruby lattice possess flat bands with finite Chern number. One can readily reach the regime $E_g/W \approx 70$.

We first study a new model without the spin-orbit coupling in Eq.\eqref{eq:H_SO}, but with the imaginary part on the hopping terms Eq.\eqref{eq:H0}. Specifically, in the Hamiltonian \eqref{eq:H0}  the hopping terms are substituted by
\begin{eqnarray}
t'&=&t+\text{i}\sigma_z t_i,\nonumber\\
t_1'&=&t_{1r}+\text{i}\sigma_z t_{1i}.
\label{eq:t1complex}
\end{eqnarray}
The complex hopping parameters above can be artificially synthesized in fermionic cold atomic optical lattices via a specific tuning of Raman fields,\cite{Osterloh:prl05, Lin:prl09, Spielman:pra09} or they can arise in some spin-orbital perovskites where strong intrinsic spin-orbit coupling can make the hopping complex and spin-dependent.\cite{Shimoyamada:prl06}

To explicitly break time-reversal symmetry, we assume the fermions are spin polarized and calculate the band structure and Chern number for one spin orientation. The Chern number of the $n$-th band is defined as\cite{Tang10}
\begin{equation}
c_n=\frac{1}{2\pi}\int_{BZ}d^2 k F_{12}(k),
\label{eq:chern}
\end{equation}
where the integral is taken for the two-dimensional Brillouin zone, and
 $F_{12}(k)$ is the associated field strength defined as
\begin{equation}
F_{12}(k)=\frac{\partial}{\partial k_1}A_2(k)-\frac{\partial}{\partial k_2}A_1(k),
\end{equation}
where $A_{\mu}(k)=-i\langle u_{n\bf k}|\frac{\partial}{\partial k_{\mu}}|u_{n\bf k}\rangle$ is
the Berry connection and $|u_{n\bf k}\rangle$ are the Bloch wavefunctions of the $n^{th}$ band.  The full Chern number is the sum of $c_n$ over all occupied bands, which in our case is just the single lowest energy band.  The calculation of Chern numbers is based on the methods in Ref.[\onlinecite{Fukui:jpsj05}], where $F_{12}(k)$ is expressed in some $U(1)$ gauge transformation link variables, and then the integration is converted into a summation in the $k$-space. With only the hopping parameters \eqref{eq:t1complex} our flattest band with finite Chern number had $W/E_g \approx 13$, which is not very large.

\begin{figure}
\centering
\includegraphics[width=0.8\linewidth]{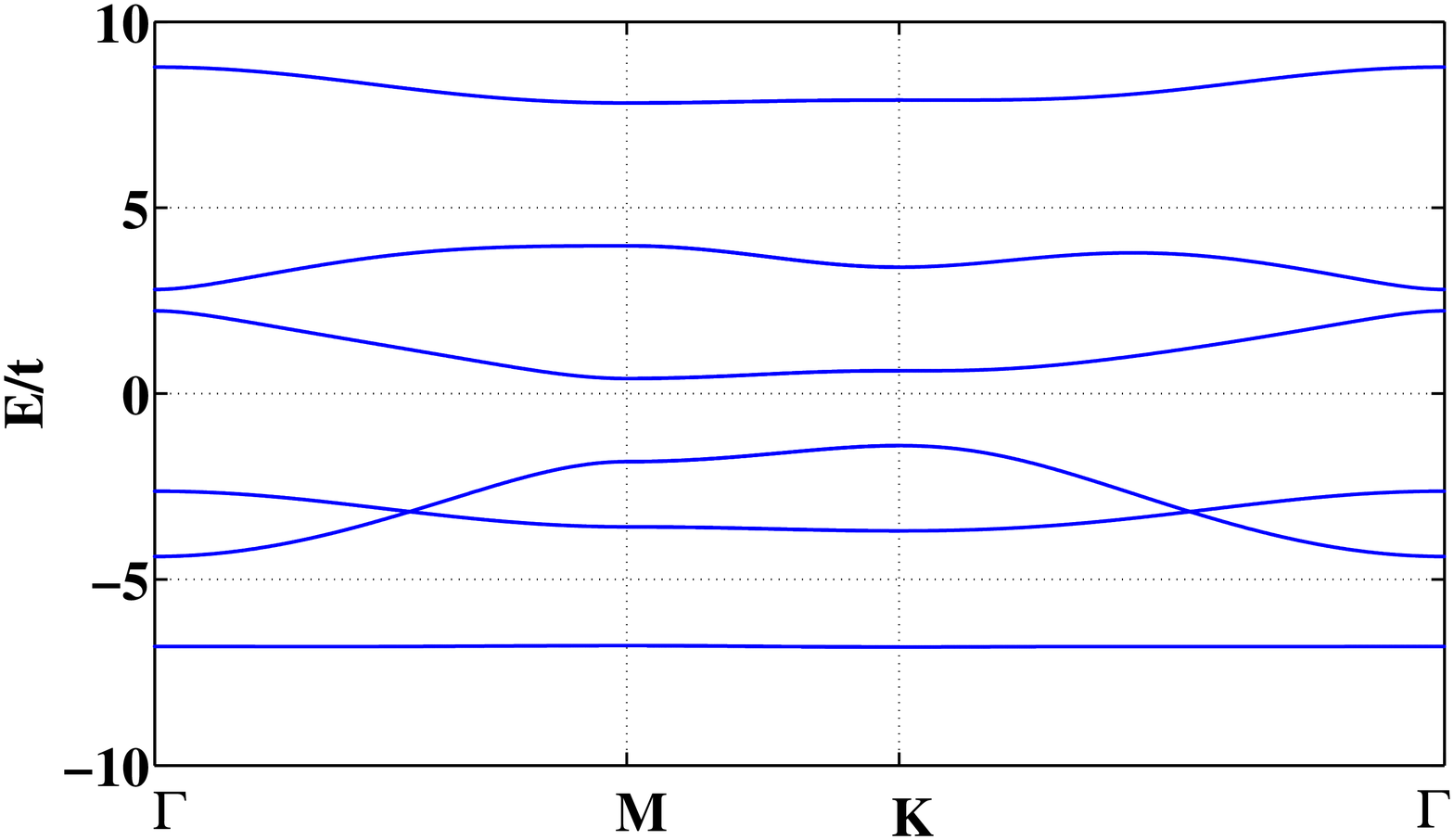}
\caption{The energy bands with $t_i=1.2t, t_{1r}=-1.2t, t_{1i}=2.6t, t_{4r}=-1.2t$.}\label{fig:t4rbands}
\end{figure}

To obtain a nearly flat band with nonzero Chern number, we needed to add more terms to the Hamiltonian. One option is to add the hopping inside the square in the diagonal directions, which is called $t_{4r}$.  This is shown schematically in Fig.~\ref{fig:ruby}(c). The nonzero Chern numbers and nearly flat bands do occur in this case. For example, with $t_i=1.2t, t_{1r}=-1.2t, t_{1i}=2.6t, t_{4r}=-1.2t$ the Chern number is -1 and the $\text{gap}=2.398t, \text{and band width}=0.037t$, which gives $W/E_g \approx 70$.  The corresponding band structure is shown in Fig.~\ref{fig:t4rbands}.  Based on the results of exact diagonalization studies on other lattices\cite{Sheng11,Neupert11,Wang11} and the general arguments given before,\cite{Sun10,Neupert11,Tang10,Wang11} we expect a flat band fractional quantum Hall effect to be realized on the ruby lattice as well for the appropriate fractional filling of the flat band.  This could form the basis of a lattice model of a fractional topological insulator\cite{Levin:prl09} by taking two time-reversed copies.\cite{Kane:prl05,Kane_2:prl05}

\section{Conclusion}
\label{sec:conclusions}

In conclusion, we have studied the energy bands of a tight-binding model on the
ruby lattice, and also obtained the phase diagrams for a variety of filling fractions. The phase diagram is rather complex in its dependence on the various types of spin-orbit coupling and nearest-neighbor hopping we considered.  We have seen how various spin-orbit coupling terms can ``compete" with each other and also ``support" each other.

In a related spinless model we calculated the Chern number for the nearly flat bands, and showed that they will likely support a robust fractional quantum Hall effect with $W/E_g \approx 70$.  This could lead to possible lattice models of a fractional topological insulator.\cite{Levin:prl09}

Since the ruby lattice has earlier appeared in the study of topological spin models,\cite{Bombin:prb09,Kargarian:njp10} the current work opens the way for an exploration of the transitions between various interesting phases in a more general interacting model.\cite{Fiete11}  Current technology in cold atomic gases should allow an experimental study\cite{Sorenson:prl05,Hafezi:pra07,Duan:prl03,Ruostekoski:prl09,Bercioux:pra09,Bercioux:pra11,Goldman:pra11} (although perhaps with challenges) of fermions in the ruby lattice and will likely raise new questions to challenge theory.

\acknowledgments
We thank Jun Wen for helpful discussions, and gratefully acknowledge
funding from ARO grant W911NF-09-1-0527 and NSF grant DMR-0955778. 
The authors acknowledge the Texas Advanced Computing Center (TACC) 
at The University of Texas at Austin for providing computing resources 
that have contributed to the research results reported within this paper. 
URL: http://www.tacc.utexas.edu


%

\end{document}